\documentclass[pra,twocolumn,superscriptaddress,showpacs]{revtex4}
\usepackage{hyperref}
\usepackage{amsmath, amssymb}
\usepackage{color}
\usepackage[titletoc]{appendix}
\usepackage{graphicx,epsfig}
\usepackage{pifont}
\usepackage{subfigure}
\usepackage{amsmath, amssymb, graphics}
\usepackage{bbold}

\begin{document}

\title{Quantum state engineering in hybrid open quantum systems}
\author{Chaitanya Joshi}
\email{chaitanya.joshi@york.ac.uk}
\affiliation{Department of Physics and York Centre for Quantum Technologies, University of York, Heslington, York, YO10 5DD, UK }
\author{Jonas Larson}
\affiliation{Department of Physics, Stockholm University, Albanova physics center, Se-106 91 Stockholm, Sweden}
\author{Timothy P. Spiller}
\affiliation{Department of Physics and York Centre for Quantum Technologies, University of York, Heslington, York, YO10 5DD, UK }
\date{\today}

\begin{abstract}
We investigate a possibility to generate  non-classical states in {\it light-matter} coupled noisy quantum systems, namely the anisotropic Rabi and Dicke models. In these hybrid quantum systems a competing influence of coherent internal dynamics and environment induced dissipation drives the system into non-equilibrium steady states (NESSs). Explicitly, for the anisotropic Rabi model the steady state is given by an incoherent mixture of two states of opposite parities, but as each parity state displays light-matter entanglement we also find that the full state is entangled. Furthermore, as a natural extension of the anisotropic Rabi model to an infinite spin subsystem, we next explored the NESS of the anisotropic Dicke model. The NESS of this linearized Dicke model is also an inseparable state of light and matter. With an aim to enrich the dynamics beyond the sustainable entanglement found for the NESS of these hybrid quantum systems, we also propose to combine an all-optical feedback strategy for quantum state protection and for establishing quantum control in these systems. Our present work further elucidates the relevance of such hybrid open quantum systems for potential applications in quantum architectures. 
\end{abstract}

\maketitle
\section{Introduction}
Light-matter coupled quantum systems are now seen as novel composite systems for the exploration of various tasks in quantum computation and information processing, as well as for testing foundations of quantum physics \cite{gkur15, azei99,jmra01}. These hybrid systems promise to combine  disparate quantum degrees of freedom in constructing  scalable quantum  architectures \cite{gkur15,dlos98}. In particular, the idea is to take `the best of two worlds': matter provides good candidates for storage of quantum information, while light (photons) are superior when it comes to transmitting or sending quantum information, with interactions between these provide processing. For example, photons can be sent between different parties in order to create entanglement over macroscopic distances. However, physical quantum systems, in this case especially the light part, invariably also couple to their external environments~\cite{uwei93,hpbr02}. It is commonly  argued that non-classical states, including entangled states, are  extremely sensitive to noise and dissipation. For example,  environment induced decoherence tends to reduce quantum-coherent superpositions to incoherent mixtures \cite{whzu03}. Quantum state engineering strategies have therefore taken a center stage in salvaging quantum coherence in hybrid  quantum systems \cite{gkur15,zeli13}. It is clear that for using photons as information carriers between different matter subsystems, sustainable entanglement between light and matter is a necessity, i.e. such quantum correlations should survive any realistic decoherence/noise affecting the photons. 

In a somewhat parallel approach, noisy coupled quantum systems have been proposed as exciting avenues for controlled generation of  multi-partite entangled states \cite{fver09}.
The interplay between coherent and incoherent dynamics in interacting  multi-partite open quantum systems can result in the generation of  steady states which exhibit  exotic quantum character \cite{cjosh13comment,cjosh13}. To take a particular example relevant for this work, for open optical systems the dynamics can often be well described by a Lindblad type master equation 
\begin{equation}\label{open}
\dot{\hat{\rho}}= -i[\hat H_{\rm S},\hat{\rho}]+\gamma\hat{\mathcal L}_{\hat A}\hat{\rho},
 \end{equation}
where $\hat\rho$ is the system's state, $\hat H_\mathrm{S}$ the system Hamiltonian, and $\hat{\mathcal L}_{\hat A}\hat{\rho}$ is the Lindblad super operator rendering the effects of the reservoir on the system ($\hat A$ is the so called quantum jump operator, and we note that $\hat H_\mathrm{S}$ may be modified from the closed system Hamiltonian due to the interaction with the environment). The steady state is solved by letting $i[\hat H_{\rm S},\hat{\rho}_\mathrm{ss}]-\gamma\hat{\mathcal L}_{\hat A}\hat{\rho}_\mathrm{ss}=0$. It follows that if $\hat\rho_\mathrm{ss}$ is a dark state, i.e. $\hat{\mathcal L}_{\hat A}\hat{\rho}_\mathrm{ss}=0$, then $\hat\rho_\mathrm{ss}$ must also commute with $\hat H_\mathrm{S}$, e.g. being an eigenstate of the Hamiltonian. However, this is not the only possibility for having a steady state; both contributions might be non-zero but also exactly cancel each other. This would mean that the unitary evolution `balances' the non-unitary evolution. For infinite systems, this interplay may result in non-equilibrium phase transitions between phases supported either by the first or the second term in Eq.(\ref{open})~\cite{cjosh13comment,cjosh13}. That is to say that $\hat\rho_\mathrm{ss}$ changes qualitatively and shows non-analytic properties at some critical $\gamma_c$. For finite systems, on the other hand, such a transition is smooth/analytic, but nevertheless for the intermediate stages between the two extremes  non-classical states may persist. In this work we explore this scenario and examine the possibility of generating non-classical states in two open quantum systems that have served as work horses in quantum optics for more than half a century, namely the Rabi~\cite{iira36} (or non-Rotating-Wave-Approximation (non-RWA) Jaynes-Cummings model~\cite{jc}) and the Dicke model~\cite{bmga11}. In a way, in our study these two models are at the two extremes of the {\it quantum spectrum}; the Rabi model which couples a single spin-1/2 particle to a boson or photon mode lives in the deep quantum regime, while what we term the Dicke model considers instead a particle with an infinite large spin $S$ (almost classical) coupled to the boson mode. As already mentioned, the main source of decoherence is very often photon absorption and we thereby consider a Lindblad term with $\hat A=\hat a$ being the photon annihilation operator. This photon loss mechanism aims at emptying the boson mode (the only dark state is the vacuum). However, for a large light-matter coupling in the Hamiltonian even the ground state (in the absence of environmental coupling) becomes a `highly excited' (from the perspective of photon number) state, and so the two dynamical contributions in the evolution of the hybrid system must be balanced in order to support a NESS. We will demonstrate that as a result of this competition between unitary and non-unitary evolution, for both models a NESS exists that possess non-classical features in terms of light-matter entanglement. Furthermore, especially for the Rabi model this NESS reflects the well known parity symmetry of the original Hamiltonian; the NESS consists of an incoherent mixture of different parity states.

In the second half of the paper we explore possibilities to systematically enrich and enhance the non-classical properties of the NESS. More precisely, although environment induced dissipation potentially can be tailored to achieve desired quantum states~\cite{fdim07}, establishing quantum control over various dissipation channels is also an important benchmark toward realizing a useful quantum architecture~\cite{hmwi10}. Thus, it is desirable to construct quantum control schemes for hybrid quantum processors if, for instance, one is interested in initial quantum state preservation~\cite{prab06}.  Along these lines we will use an all-optical coherent feedback strategy~\cite{hmwi94,ptom01,cjosnj14} to achieve quantum state preservation. It is important to appreciate that such a scheme is different from just tailoring the dissipation channels as it involves active feedback to the system. As outlined in Refs.~\cite{hmwi94,ptom01}, under the assumption that the time-delay introduced by the feedback loop is negligible it is indeed possible to establish a complete control over the dissipative dynamics. Going beyond such idealized sudden feedback control, we will conclude by briefly commenting  on the prospects of our scheme by  combining it with a  time-delayed feedback control scheme as previously studied  both in the classical \cite{kpyrg92} and quantum domains \cite{nyam14,algr14}.

The paper is organized as follows. In Sec.\ref{sec:model1} we introduce our first physical  model describing the interaction between a quantized bosonic mode and a single two-level system. We explore the general structure of the NESS $\hat\rho_\mathrm{ss}$ (being of an interesting form) and extract the sustainable entanglement between the two subsystems. The following Sec.\ref{sec:model2} addresses the question whether non-classical features survive in the `classical limit' in which the two-level system is turned into an infinite one {\it i.e.} identifying the linearized Dicke model. And indeed, entanglement is sustained also in this case. Having discussed these two model examples, in Sec.~\ref{sec:feedbck} we combine a coherent all-optical feedback scheme for quantum state protection in open quantum systems. While entanglement is sustainable in the original open systems, the feedback scheme allows for increasing the desired properties. Finally, we conclude the paper with a short  discussion in Sec.~\ref{sec:concld}.

\section{model A - Anisotropic Rabi model}  \label{sec:model1} The quantum Rabi model describes  the interaction between a quantized bosonic field and the simplest quantum mechanical model, a single  two-level system or qubit \cite{iira36}. Despite its simplicity, the quantum Rabi model had eluded an exact solution  for many decades and it was only recently that an analytical solution was obtained~\cite{dbra11}, with continuing debate as to the completeness of this solution and the existence of other possible solutions~\cite{mor}.  Nonetheless, over the years the quantum Rabi model has turned out to be an ubiquitous physical model and has found applications in understanding a wide variety of physical systems, including  trapped ions, superconducting qubits, optical and microwave cavity QED, circuit QED, among others \cite{commentonRabi}. In some cases the two-level system may be an exact description of the matter part of the hybrid system (when it is indeed a spin-1/2 particle), but in many cases the two-level description is an approximation of a more complicated matter sub-system.

In its most simplified form the quantum Rabi model takes the form (with $\hbar$=1 and neglecting the vacuum energy of the boson mode)
\begin{equation}\label{rabiorig}
\hat H_\mathrm{RM}=\omega \hat{a}^{\dagger}\hat{a}+\frac{\Omega}{2} \hat{\sigma}_{z} +\lambda(\hat{a}^{\dagger}+\hat{a})(\hat\sigma^{+}+\hat\sigma^{-}).
\end{equation}
Here, $\hat{a}$ and $\hat{a}^{\dagger}$ are the annihilation and creation operators for the bosonic field of frequency $\omega$, $\hat\sigma^{\pm}=(\hat\sigma_{x}\pm i\hat\sigma_{y})/2$ with $\hat\sigma_{x,y,z}$ the Pauli matrices for the two-level system, $\Omega$ is  the  energy level splitting between the two levels, and the coupling strength  between the bosonic mode and the two-level system is represented as  $\lambda$ ($\geq0$). Throughout we will use dimensionless parameters such that we scale all frequencies by $\omega$ and time by $\omega^{-1}$. Nevertheless, we keep $\omega$ in all expressions in order to keep track of all terms and instead always use $\omega=1$ in any calculations. The counter-rotating terms in the Hamiltonian \eqref{rabiorig}, $\hat{a}^{\dagger}\hat\sigma^{+}$ and $\hat\sigma^{-}\hat{a}$, do not conserve the excitation number. However, if the coupling $\lambda \ll\omega,\Omega$, these counter-rotating terms can be dropped from the Hamiltonian \eqref{rabiorig} under the so-called Rotating Wave Approximation (RWA)~\cite{jc,dbal12}.  Simplifying the  Hamiltonian \eqref{rabiorig} under the RWA results in the widely studied Jaynes-Cummings model which can be straightforwardly solved \cite{jc,bwsh93}. As evident through several standard cavity-QED experiments,  the atom-field coupling $\lambda$ is normally orders of magnitude smaller than the bare transition frequencies $\omega,\Omega$,  and this makes the RWA a valid approximation to simplify  the quantum Rabi model \cite{jmra01}. The straightforward solvable nature of the Jaynes-Cummings model results from the presence of a continuous $U(1)$ symmetry; the total number of excitations $\hat N=\hat a^\dagger\hat a+\frac{1}{2}\hat\sigma_z$ is preserved. This should be contrasted with the discrete $Z_2$ parity symmetry of the Rabi model characterized by the unitary $\hat U_{Z_2}=\exp\left[i\left(\hat a^\dagger\hat a+\hat\sigma^+\hat\sigma^-\right)\pi\right]$. The action of $\hat U_{Z_2}$ is to flip the signs of $\hat a$, $\hat a^\dagger$, $\hat\sigma_{x,y}$ while leaving $\hat\sigma_z$ invariant. Thus, the Rabi model has a lower symmetry than the Jaynes-Cummings model, which naturally is the reason why finding an exact Rabi solution is very hard.

The RWA simplified quantum Rabi model is undoubtedly one of the most celebrated models in quantum optics and has received  well deserved attention \cite{bwsh93, dlei03, jlar07}. However, there has been a recent surge of interest in exploring hybrid quantum systems  in the so-called ``ultra-strong'' coupling regime \cite{tnie10}. In this ultra-strong coupling regime the atom-field coupling strength $\lambda$  can approach a non-negligible fraction of the bare transition frequencies $\omega,\Omega$, thereby making the Jaynes-Cummings model a non-valid approximation for the quantum Rabi model \cite{jqyo11}. From the discussion of the introduction, it should be clear that in order to generate large steady state entanglement the coupling $\lambda$ should be made as large as possible (allowed by other approximations like the two-level approximation). As $\lambda>\sqrt{\omega\Omega}$ the ground state of the Rabi model undergoes a qualitative change~\cite{jlar07,irish} where the field builds up a large non-zero photon number. In the Dicke model this marks the normal-superradiant phase transition~\cite{hepp}. This extremely large $\lambda$ regime has been termed the ``deep strong'' coupling regime~\cite{deep}. Thus, as the ultra-strong coupling regime has recently been realized in certain (non-driven) circuit QED architectures~\cite{tnie10} ($\lambda\sim\omega/10$), we would like to explore the quantum Rabi model further, into the deep strong coupling regime where the coupling parameter $\lambda \sim \omega,\Omega$. It is still unclear whether or not a non-driven system could attain such couplings~\cite{commntdoma2}. Equally important, in this  deep strong coupling regime the no-go theorem tells us that the `self-energy' of the field cannot be neglected and by including such a term the passing to a large photon populated ground state will not occur~\cite{nogo}. Therefore, to overcome such hindrances we choose to work with an effective realization of the quantum Rabi model with suitably engineered Raman driving \cite{fdim07,dbal12,algr13}. Furthermore, as such driven models may derive more generalized versions of $\hat H_\mathrm{RM}$, we will work with the anisotropic Rabi model (ARM)~\cite{qtxi14}
\begin{equation}\label{rabianiso}
\hat H_{\rm ARM}=\omega \hat{a}^{\dagger}\hat{a}+\frac{\Omega}{2} \hat{\sigma}_{z} +\lambda_1\!\left(\hat{a}^{\dagger}\hat{\sigma}^{-}\!+\hat{\sigma}^{+}\hat{a}\right)+\lambda_2\!\left(\hat{a}^{\dagger}\hat{\sigma}^{+}\!+\hat{\sigma}^{-}\hat{a}\right)\!.
\end{equation}
The possibility to realize the above ARM and also to tune $\omega,\Omega,\lambda_1,\lambda_2$ relative to each other derives from the fact that amplitude of the Raman drive effectively determines the coupling parameters~\cite{fdim07,dbal12,algr13}. With this model it is also easy to extract the importance of the counter rotating terms neglected in the RWA. In particular, the above Hamiltonian reduces to the Jaynes-Cummings model when $\lambda_2=0$, and to the quantum Rabi-model when $\lambda_1=\lambda_2=\lambda$. One important observation is that the $Z_2$ parity symmetry is preserved for the anisotropic Rabi model which naturally implies that the eigenstates can be assigned an even or odd parity, i.e. simultaneous eigenstates of $\hat U_{Z_2}$ with $\pm1$ eigenvalues. In passing we may note that recently, the anisotropic Rabi model was also termed the $U(1)/Z_2$ Rabi model as it includes both the Jaynes-Cummings and Rabi models as limiting cases~\cite{anrab}. A general state (not an eigenstate) with even$(+)$/odd$(-)$ parity takes the respective forms
\begin{equation}\label{symstates}
\begin{array}{c}
\displaystyle{|\Psi_+\rangle=\sum_{n}\left(c_{n}^\uparrow|2n+1\rangle|\uparrow\rangle+c_{n}^\downarrow|2n\rangle|\downarrow\rangle\right)},\\ \\
\displaystyle{|\Psi_-\rangle=\sum_{n}\left(d_{n}^\uparrow|2n\rangle|\uparrow\rangle+d_{n}^\downarrow|2n+1\rangle|\downarrow\rangle\right)},
\end{array}
\end{equation}
where the first ket state $|n\rangle$ represents the boson Fock state and the second $|\!\!\uparrow\rangle$/$|\!\!\downarrow\rangle$ the ``up/down'' eigenstate of the Pauli $z$ matrix. Naturally, the coefficients $c_{n}^{\uparrow,\downarrow}$ and $d_{n}^{\uparrow,\downarrow}$ will depend on the system parameters and, due to the simplicity of the Rabi model, they can in principle be accurately obtained by numerical diagonalization of \eqref{rabianiso}. Let us give one more comment related to the symmetries of $\hat H_\mathrm{ARM}$. To make this comment clear we assume that the boson mode is only weakly populated such that we can truncate it to contain only the vacuum and the first Fock state. In this case we relabel the creation/annihilation operators with $\hat\tau^+/\hat\tau^-$ respectively. In this truncated space we can define effective Pauli operators $\hat\tau^+=(\hat\tau_x+i\hat\tau_y)/2$, $\hat\tau^-=(\hat\tau_x-i\hat\tau_y)/2$, and $\hat\tau_z=2\hat\tau^+\hat\tau^--1$. Within this approximation and formulation we rewrite the Hamiltonian as (up to a constant)
\begin{equation}\label{rabianiso2}
\hat H_{\rm ARM}=\omega \hat\tau_z+\frac{\Omega}{2} \hat{\sigma}_{z} +\frac{\lambda_1+\lambda_2}{2}\hat\tau_x\hat\sigma_x++\frac{\lambda_1-\lambda_2}{2}\hat\tau_y\hat\sigma_y.
\end{equation}
Expressed in this form we can identify the Jaynes-Cummings model ($\lambda_2=0$) with the two site Heisenberg $XX$ model and the anti-Jaynes-Cummings model~\cite{ajc} ($\lambda_1=0$) also with the $XX$ model but with one coupling supporting ferro- and the other anti-ferromagnetic order. Similarly, the Rabi model ($\lambda_1=\lambda_2$) can be identified with the two site transverse Ising model, and finally the ARM  ($\lambda_1\neq\lambda_2$) with the two site Heisenberg $XY$ model. Naturally, the character of the couplings are not just limited to the above approximation where the boson Hilbert space has been truncated to two states, but holds also for the original Hamiltonian.  We cannot talk about phase transitions in these finite size models (contrary to the Dicke model), but nevertheless the types of coupling share great similarities with these spin systems which all are critical in the thermodynamic limit~\cite{jj}. We note that the Dicke phase transition is within the Ising model universality class. Even though we cannot properly take a thermodynamic limit of the Rabi model, it was shown that typical features (non-analyticity and vanishing order parameter in one phase) of a phase transition can also be achieved in the Rabi model by letting $\Omega/\omega\rightarrow\infty$. Not surprisingly, here one also finds universal Ising behavior~\cite{plenio}.

In \cite{fdim07} an effective optical realization of the  Hamiltonian \eqref{rabianiso} based on multi-level atoms and cavity-mediated Raman transitions has been presented. Most recently this scheme was also experimentally demonstrated, and importantly the deep strong coupling regime of the Dicke model was reached~\cite{dexp1}. A many-body generalization of the Hamiltonian \eqref{rabianiso}  has also been recently considered, to explore the non-equilibrium phase diagram of the dissipative Rabi-Hubbard model \cite{msch15}. We will not enter into the details of a physical realization of the Hamiltonian \eqref{rabianiso}, but will refer the reader to a comprehensive analysis performed in  \cite{fdim07}.

Following \cite{fdim07}, damping of the field mode is the dominant source of dissipation and so we will approximate by taking it as the only  dissipation channel in the open version of the ARM \eqref{rabianiso}. This assumption is particularly valid for the scheme outlined in \cite{fdim07} where the metastable  low lying energy doublet of a Raman driven four level atom can act as a qubit in the ARM \eqref{rabianiso}. Under  the Born-Markov and secular approximations~\cite{uwei93,hpbr02}, we model the evolution of the  joint atom-field density operator $\hat\rho$ by the master equation 
\begin{equation}\label{rabianisoopen}
\dot{\hat{\rho}}= -i[\hat H_{\rm ARM},\hat{\rho}]+\gamma\hat{\mathcal L}_{\hat{a}}\hat{\rho},
 \end{equation}
 where $\gamma$ is the damping rate of the field and $\hat{\mathcal L}_{\hat{x}}\hat{\rho}=2\hat{x}\hat{\rho}\hat{x}^{\dagger}-\hat{x}^{\dagger}\hat{x}\hat{\rho}-\hat{\rho}\hat{x}^{\dagger}\hat{x}$ is the Lindblad super operator.  It is worth noting that in writing the above master equation we  have assumed that the cavity damping is independent of the inter-mode coupling strengths $\lambda_1$ and $\lambda_2$. The form of the master equation \eqref{rabianisoopen} can only  arise out of an original  time-dependent (driven) Hamiltonian.  In the approach outlined in \cite{fdim07}  the coupling Hamiltonian \eqref{rabianiso} is written in the frame of an external drive, imparting  it a non-equilibrium character. For a time-independent system coupled to an external bath, it is required that the steady state will obey the principles of  equilibrium statistical mechanics \cite{cjos14}. No such restriction is required on the dynamics arising out of a implicit time-dependent Hamiltonian.  Naturally, this is crucial in our study, as we reach steady states that are different from a thermal equilibrium state. We thereby can conclude that even if deep strong coupling regime could be reached without external pumping such a model would be conceptually different from the present one as the Lindblad jump operators would be different~\cite{blais}.

In order to analyze the particular structure of the steady states $\hat\rho_\mathrm{ss}$ we numerically solve the master equation \eqref{rabianisoopen} and explore the resulting NESS. The steady state is found by integrating the equation for long times, for various initial states, and checking for convergence of the solution. According to the above discussion, a steady state cannot be a dark state in this case since we know that the dark state is the vacuum which in return is not an eigenstate of the Hamiltonian. 
We further know that since the system is driven, the NESS should be different from the system ground state which becomes the steady state when the system is coupled to a zero temperature bath~\cite{blais}. However, less clear is whether the steady state is unique. This could, in principle, be checked by diagonalizing the master equation, but here we only remark that our numerical simulations, for given parameters and for different initial states, all relax to steady states of a specific form.

Through numerical evaluation of the master equation  \eqref{rabianisoopen} we arrive at the atom-field  NESS density matrix $\hat\rho_{\rm ss}$ and find it to have an approximate  structure when ordering the elements in a specific way. In a three-excitation manifold, for instance, $\hat\rho_{\rm ss}$  can be explicitly expressed as
\begin{equation}\label{structdens}
\resizebox{.95\hsize}{!}{$\hat\rho_{\rm ss}=\bordermatrix{\text{}& \langle 0 \uparrow|&\langle 0 \downarrow|&\langle 1 \uparrow| & \langle1 \downarrow| &\langle 2 \uparrow| & \langle2 \downarrow|& \langle 3 \uparrow| & \langle3 \downarrow| \cr
                |0 \uparrow\rangle&\times&&&\times &\times&&&\times&\cr
                 |0 \downarrow\rangle&&\times&\times &&&\times&\times&&\cr
                 |1 \uparrow\rangle&&\times&\times&&&\times&\times&\cr
                |1 \downarrow\rangle&\times&&& \times&\times&&&\times\cr
                 |2 \uparrow\rangle&\times&&&\times&\times&&&\times\cr
                |2\downarrow\rangle&&\times&\times&&&\times&\times&&\cr
                  |3 \uparrow\rangle&&\times&\times&&&\times &\times&\cr
                |3\downarrow\rangle&\times&&&\times&\times&&&\times\cr}$ },
\end{equation}
where the crosses mark the non-vanishing elements. The above structure tells us that  the overlap between the states of opposite parities in the  steady state density matrix $\hat\rho_\mathrm{ss}$ is identically zero.

We can give a qualitative argument which allows the above structure of the density matrix $\hat\rho_{\rm ss}$. The Lindblad super operator is obviously invariant under the action of $\hat U_{Z_2}$; $\hat{\mathcal L}_{\hat a}\hat\rho=\hat{\mathcal L}_{\hat U_{Z_2}\hat a\hat U_{Z_2}^{-1}}\hat\rho$. As discussed in detail above, the ARM \eqref{rabianiso} is also invariant under the action of parity operator $\hat U_{Z_2}$. Combining these two observations might tempt us to believe that the entire Lindblad master equation (\ref{rabianisoopen}) is also symmetric under $\hat U_{Z_2}$. However, it is important  to recognize that for open quantum systems a symmetry does not necessarily imply a conserved quantity~\cite{symcon}. Thus, in general we do not have such a thing as Noether's theorem for open quantum systems. The numerically obtained  steady state density matrix  $\hat\rho_\mathrm{ss}$ confirms this  conjecture. In other words, the parity conservation of the original ARM \eqref{rabianiso} is broken under the evolution described by the Lindblad master equation \eqref{rabianisoopen}.

The steady state density matrix $\hat\rho_\mathrm{ss}$  \eqref{structdens} can also be expressed in a closed form as 
\begin{equation}\label{stdystateaqrm2}
\hat\rho_\mathrm{ss}=\cos^2\theta|\Psi_+\rangle\langle\Psi_+|+\sin^2\theta|\Psi_-\rangle\langle\Psi_-|,
\end{equation}
with the parity states (\ref{symstates}), and $\theta$ is some constant determined by the system parameters. Thus, any coherence between the states of opposite parities is lost and the steady state is an incoherent mixture of these. Nevertheless, in the steady state quantum coherence can survive among the states with definite parity. The parameter $\theta$ is in general not $0$ nor $\pi$, meaning that parity is not a conserved quantity under evolution of (\ref{rabianisoopen}); for example, an initial state with a definite  parity will typically also end up in an incoherent mixture of the two parity states $|\Psi_\pm\rangle$. In the language of `pointer states'~\cite{point}, the states robust to the present decoherence are those with certain parity.

 In support of  the structure of the steady state density matrix \eqref{stdystateaqrm2} we argue that it is due to a symmetry of the equation of motion \eqref{rabianisoopen}, $\hat{\rho}\rightarrow \hat U_{Z_2} \hat{\rho} \hat U_{Z_2}^{-1}$. Let us assume that the steady state of the master equation \eqref{rabianisoopen} has off-diagonal coherence present between the sectors of opposite parities and has a structure 
\begin{eqnarray}
\hat\rho_\mathrm{ss}=\cos^2\theta|\Psi_+\rangle\langle\Psi_+|+\sin^2\theta|\Psi_-\rangle\langle\Psi_-|\nonumber \\
+\delta|\Psi_+\rangle\langle\Psi_-|+\delta^{*}|\Psi_-\rangle\langle\Psi_+|,
\end{eqnarray}
where $\delta$, likewise $\theta$, is some constant determined by the system parameters. If  the above $\hat\rho_\mathrm{ss}$ is a steady state of the master equation \eqref{rabianisoopen} then it should respect the above symmetry {\it i.e.} 
\begin{equation}
\hat U_{Z_2} \hat\rho_\mathrm{ss} \hat U_{Z_2}^{-1}\nonumber=\hat\rho_\mathrm{ss}.
\end{equation}
Since $\hat U_{Z_2}|\Psi_\pm\rangle=\pm|\Psi_\pm\rangle$, we get 
\begin{equation*}
\delta|\Psi_+\rangle\langle\Psi_-|+\delta^{*}|\Psi_-\rangle\langle\Psi_+|=0,
\end{equation*}
and we recover the above structure of the steady state density matrix \eqref{stdystateaqrm2}.  It is worth pointing out that a two-site steady state density matrix of the dissipative transverse field Ising model can be deduced as a special case of our density matrix structures \eqref{structdens},\eqref{stdystateaqrm2} \cite{cjosh13}. This is not a surprise since both the Rabi and the transverse field Ising models have discrete  symmetries  and which are also obeyed by the respective equations of motion.

\begin{figure}[h!]
  \centering
    \includegraphics[width=0.48\textwidth]{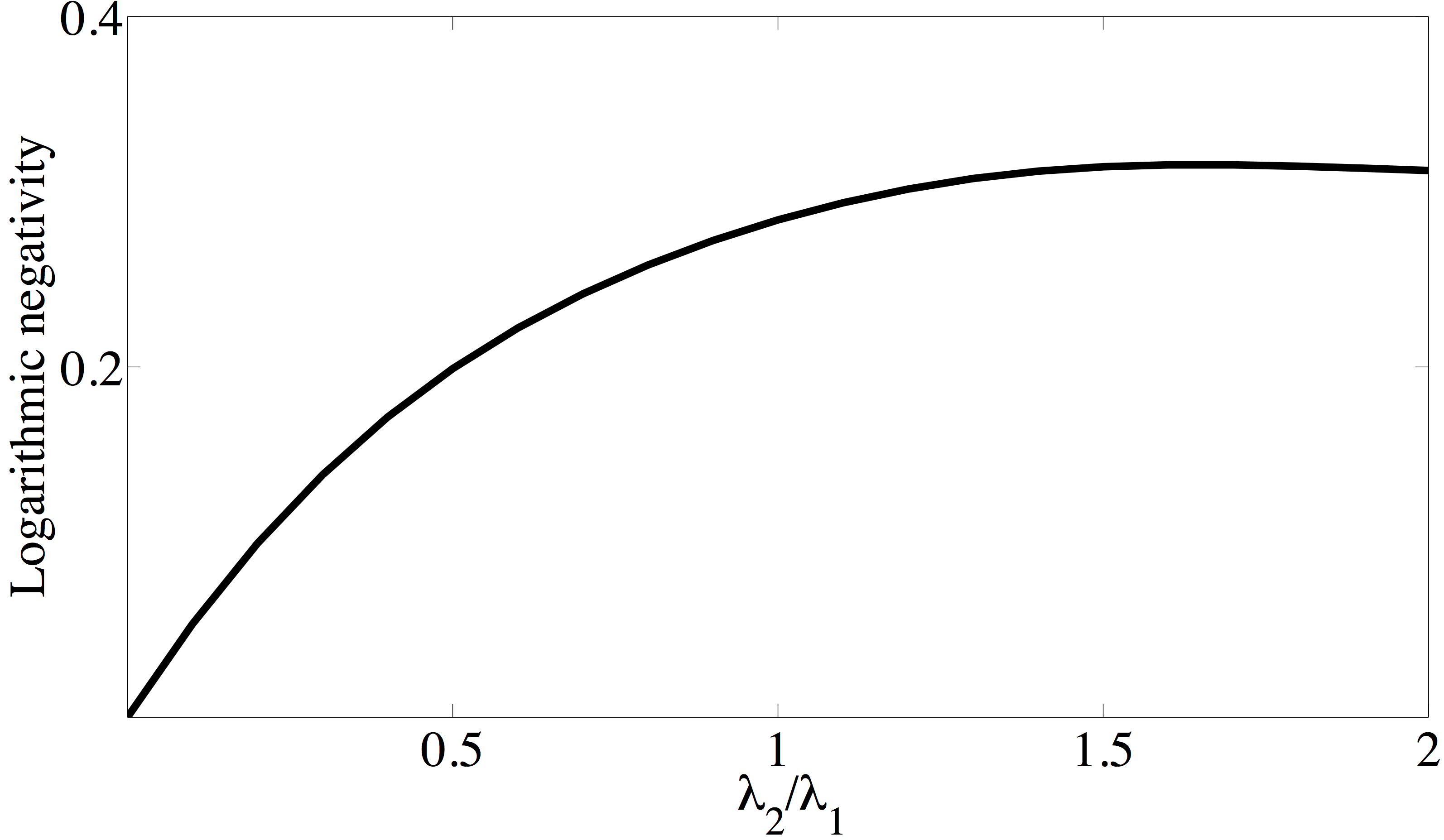}
    \caption{(Color online) Steady state entanglement (logarithmic negativity)  in the joint atom-field steady state $\hat\rho_{\rm ss}$, plotted here as a function of the dimensionless coupling ratio $\lambda_2/\lambda_1$. In the limit of vanishing $\lambda_1$ or $\lambda_2$ the steady state is separable. The maximum entanglement obtained for $\lambda_2\approx1.6\lambda_1$ demonstrates that the presence of photon losses break the symmetry between the Jaynes-Cummings and anti-Jaynes-Cummings models. In particular, the counter rotating terms are responsible for counteracting the losses. The other dimensionless parameters are $\Omega=\omega=1$ and $\gamma=0.1$, and for the plot we fix $\lambda_1=0.5$.}
\label{cavatment}
\end{figure}

As seen from from Eq.~(\ref{stdystateaqrm2}), the presence of the driving together with the reservoir demolishes any quantum coherence between the two different parity components. Nevertheless, the steady state will in general be an inseparable state of the atom and the field. To quantify the amount of this bi-partite entanglement present in the atom-field joint density matrix we compute the logarithmic negativity defined as ${\rm log} (\sum_{i}^{n}|\Xi_{i}|)$, where $\sum_{i}^{n}|\Xi_{i}|$ is the sum of the absolute values of all the eigenvalues of the partially transposed density matrix \cite{aper96,gvid02}. A non-zero value of the logarithmic negativity is sufficient to ensure inseparability of the steady state density matrix $\hat\rho_{\rm ss}$. The steady state logarithmic negativity, obtained from numerical integration of the master equation to first obtain $\hat\rho_\mathrm{ss}$ and then partially transpose it, is plotted as a function of the dimensionless coupling ratio $\lambda_2/\lambda_1$ in Fig.~\ref{cavatment}. As can be seen, the steady state entanglement between the atom and the field grows with the value of $\lambda_2/\lambda_1$. In the limit $\lambda_2/\lambda_1=0$ the case of the Jaynes-Cummings model is recovered and we know that the steady state is simply $\hat\rho_\mathrm{ss}=|0\rangle\langle0|\otimes|\downarrow\rangle\langle\downarrow|$. Thus, the counter rotating terms are responsible for the build-up of photonic and atomic excitations and thereby also for bi-partite entanglement in the system. However, in the opposite limit, $\lambda_1/\lambda_2=0$, we obtain the anti-Jaynes-Cummings model and here the steady state is again separable; $\hat\rho_\mathrm{ss}=|0\rangle\langle0|\otimes|\uparrow\rangle\langle\uparrow|$. A maximum of the entanglement should therefore be given for some finite $\lambda_2/\lambda_1$, and from Fig.~\ref{cavatment} we see that it happens at $\lambda_2/\lambda_1\approx1.6$, i.e. it is favourable to support a stronger coupling $\lambda_2$ in order to achieve a non-classical entangled state. Even though the Jaynes-Cummings and anti-Jaynes-Cummings models are mathematically equivalent (in the bare basis, for example, the Hamiltonian maintains  a block structure with $2\times2$ blocks apart from the ground state), we see that the presence of the Lindblad term $\hat{\mathcal{L}}_{\hat{a}}$ alters this symmetry, i.e. the maximum entanglement is not obtained for $\lambda_1=\lambda_2$.  

\begin{figure}[h!]
  \centering
  \includegraphics[width=0.48\textwidth]{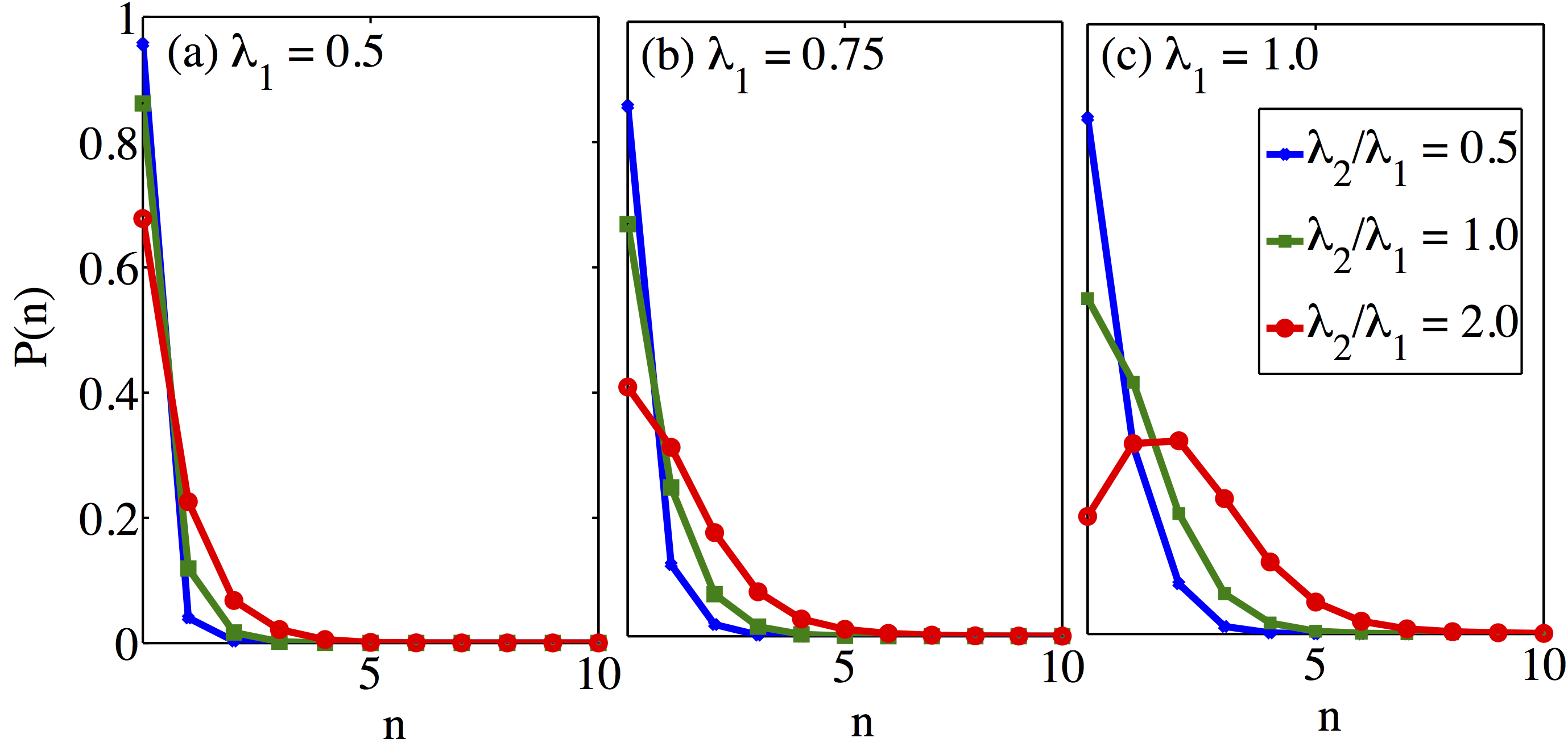}
        \caption{(Color online) Steady state photon probability distribution $P(n)$ for different coupling strengths $\lambda_1$ and $\lambda_2$. It is noted that the distribution is found to be super-Poissonian in all cases, i.e. the photon distributions alone does not display non-classical features. The lines between the points are for guiding the eye, and the dimensionless parameters are the same as in Fig.~\ref{cavatment}; $\Omega= \omega=1$ and $\gamma=0.1$.        }\label{proba_all_ratios}
\end{figure}

\begin{figure}[h!]
  \centering
  \includegraphics[width=0.5\textwidth]{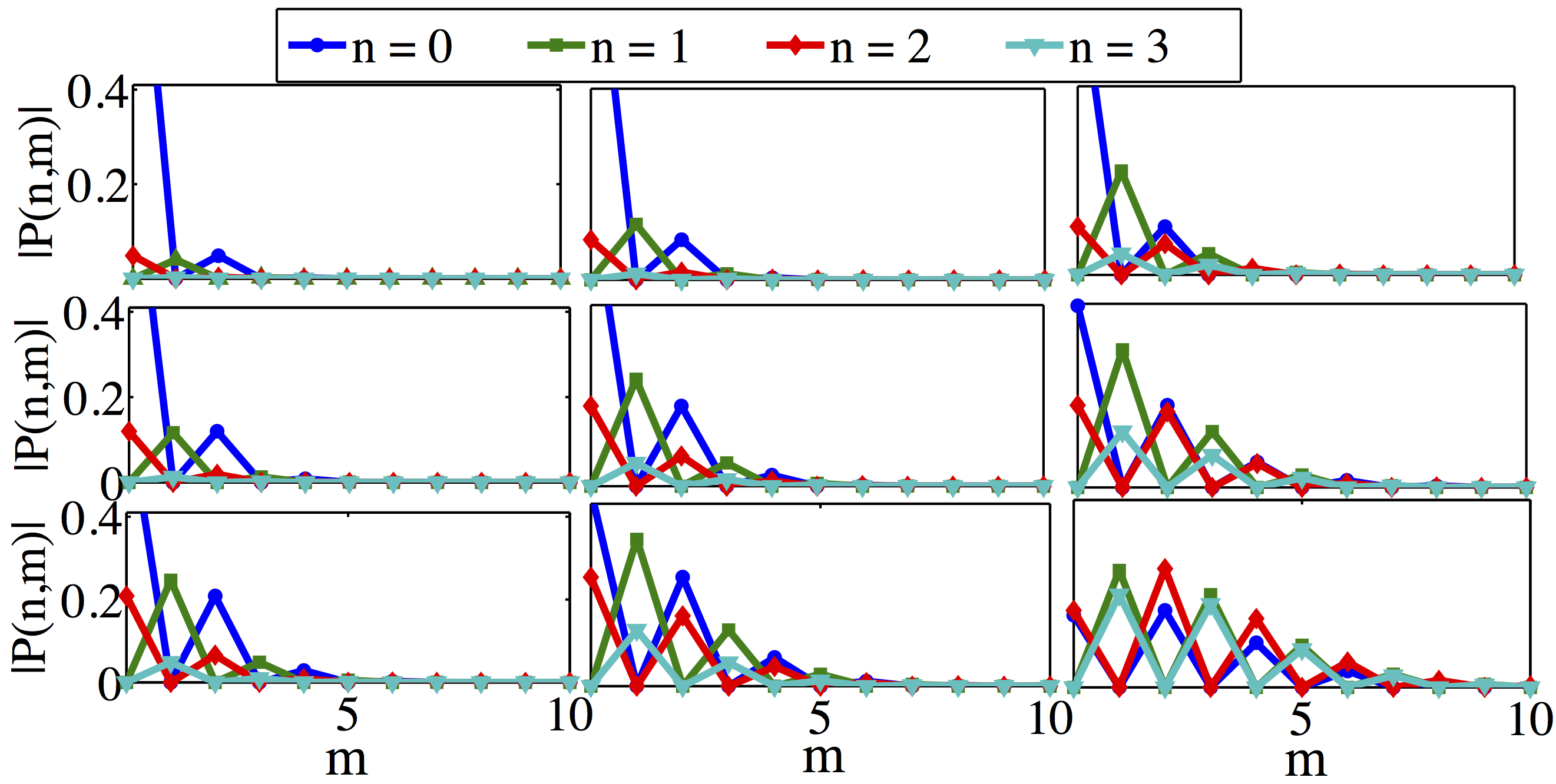}
        \caption{(Color online) Leading off-diagonal elements of the reduced density matrix $\hat\rho_{\rm field}$ \eqref{stdystatecavtyaqrm}, when $\lambda_2/\lambda_1=0.5$ (top row), $\lambda_2/\lambda_1=1.0$ (middle row), and $\lambda_2/\lambda_1=2.0$ (bottom row), and $\lambda_{1}=0.5,0.75,1.0$ (increasing from left to right in all three rows). The alternating zero and non-zero elements reflects the population of states with different parities. The other parameters are the same as for Fig.~\ref{proba_all_ratios}.}\label{coher_all_ratios}
\end{figure}

The reduced density matrix of  the boson field is obtained by tracing over the atomic degrees of freedom, $\hat\rho_{\rm field}={\rm Tr_{atom}}\left[ \hat\rho_{\rm ss}\right]$, giving
\begin{eqnarray}\label{stdystatecavtyaqrm}
\hat\rho_{\rm field}&=&\mathop{\sum_{n=0}^{\infty}\sum_{m=n,n+2...}^{\infty}} P(n,m) |n\rangle \langle m|  +h.c.~.
\end{eqnarray}
As a result of the mixture (\ref{stdystateaqrm2}), the steady state field distribution is an individual coherent mixture of even and odd numbered Fock states. However, these two coherent mixtures exist in two different `sectors' with no coherent overlap between them. In Fig.~\ref{proba_all_ratios} we plot the steady state  probability distribution $P(n)=\langle n|\hat\rho_{\rm field}|n\rangle$ for three different values of the ratio $\lambda_2/\lambda_1$, and again the steady state has been found by integration of Eq.~(\ref{rabianisoopen}). We find that the steady state density matrix  $\hat\rho_{\rm field}$  follows super-Poissonian statistics where the difference between the variance and the mean of the photon number distribution increases with the value of the ratio $\lambda_{2}/\lambda_{1}$. In particular, and as expected, for increasing atom-field coupling the mean number of photons increases. The counter rotating terms play a more important role for this increase to occur. To explore the importance of the bare state coherences in Fig.~\ref{coher_all_ratios} we  plot the absolute value of the first four leading order off-diagonal components of  $\hat\rho_{\rm field}$, namely $|P(n,m)|=|\langle n|\hat\rho_{\rm field}|m\rangle|$. It is clear that these imprint the coherence  among the bare states belonging to either even or odd parity sectors.  

\begin{figure}[h!]
  \centering
  \includegraphics[width=0.5\textwidth]{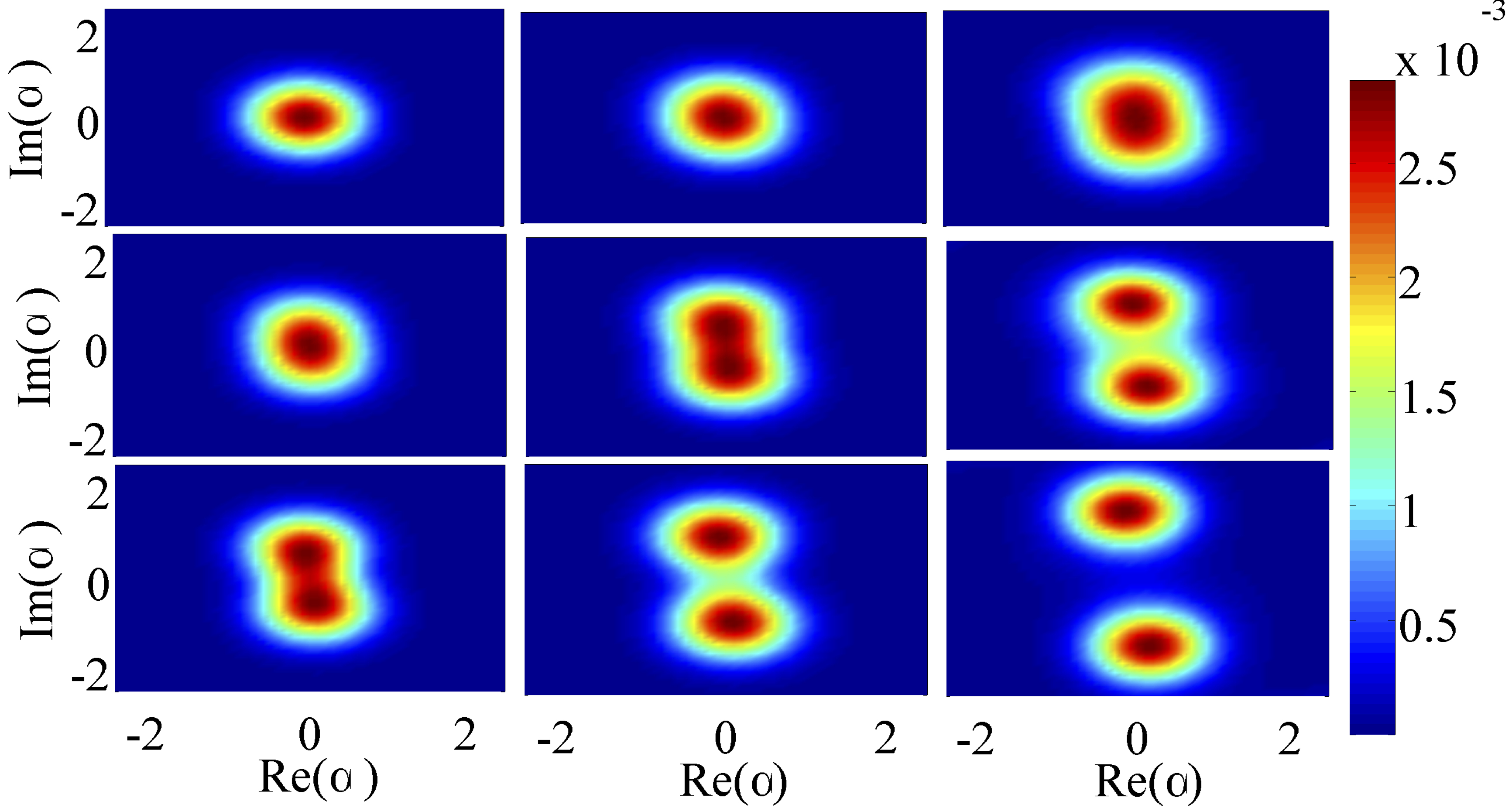}
        \caption{(Color online) Steady state Wigner function \eqref{wgner} for the anisotropic Rabi model; $\lambda_2/\lambda_1=0.5$ (top row), Rabi model $\lambda_2/\lambda_1=1.0$ (middle row), and $\lambda_2/\lambda_1=2.0$ (bottom row), and $\lambda_1=0.5,0.75,1.0$ (increasing from left to right in all three rows). The splitting between the two blobs scale as $\lambda_1+\lambda_2$ for the closed case. The presence of photon loses tend to make the splitting smaller since the corresponding Lindblad term favor a vacuum state. The bath also brings about an energy (Lamb) shift which is manifested in the tilting of the Wigner function. The absence of negativity of the Wigner functions in this plot does not solely result from the reservoir induced decoherence but also from the entanglement shared between the atom and the field. The other parameters are the same as for Fig.~\ref{proba_all_ratios}.}\label{Wigner_all_ratios}
\end{figure}

So far we have seen that for the driven/dissipative Rabi model non-classical features may survive.
An interesting question then raises, which of the features of the (equilibrium) ground state of the Rabi model \eqref{rabianisoopen} can be reproduced in  the (non-equilibrium) steady state of the master equation \eqref{rabianisoopen}? For the Rabi model, in the deep strong coupling regime the ground state becomes a Schr\"odinger cat~\cite{jlar07,irish}. The more resolved the cat state becomes (the smaller the overlap between the two separated field states in phase space becomes), the higher the atom-field entanglement obtained.  More precisely, these `dead' and `alive' field states measure the dipole moment of the atom (corresponding to eigenstates of $\hat\sigma_x$). Thus, when tracing out the atom, the field state will become a statistical mixture between `dead' and `alive'. Note that pure cat state of the form $|\mathrm{CAT}_\pm\rangle\propto(|\alpha\rangle\pm|-\alpha\rangle)$, for coherent states $|\pm\alpha\rangle$ with $|\alpha|>0$, have a definite parity and the corresponding photon distributions contain either only even or odd photon states $|n\rangle$. Thus, we ask if we can expect a  similar statistical cat also for the reduced density matrix for the boson field  \eqref{stdystatecavtyaqrm}? Figure~\ref{proba_all_ratios} reveals at least that the photon distribution of the steady state is super-Poissonian, 
and combining this with Fig.~\ref{coher_all_ratios} one may expect the cat structure to survive photon decay. This should be contrasted with the pioneering experiments in the {\it ENS} Paris group which measured the decay of the cat into a fully separable atom-field state before all photons had leaked the cavity~\cite{harochecat}. The phase space distribution of the steady state field distribution for the Rabi model is indeed split in two as is shown in Fig.~\ref{Wigner_all_ratios}. The Wigner function for the field mode can be obtained from the following transformation of the density operator
\begin{equation}\label{wgner}
W(\alpha)=\frac{2}{\pi}{\rm Tr}[\hat{D}^{\dagger}(\alpha)\rho_{\rm field}\hat{D}(\alpha)(-1)^{\hat{a}^{\dagger}\hat{a}}],
\end{equation}
where $\hat{D}(\alpha)={\rm exp}(\alpha  \hat{a}^{\dagger}-\alpha^{*}\hat{a})$ is the displacement operator and $\alpha=\alpha_{r}+i\alpha_{m}$ is a complex parameter \cite{smba97}.  Note that for other expressions for the Wigner function, the real and imaginary parts of $\alpha$ are related to `momentum' $p$ and `position' $x$.  As can be seen from the figure, on increasing the value of the ratio $\lambda_2/\lambda_1$ the separation between the two-lobes becomes  more pronounced and eventually the ``two-lobes'' cease to overlap with each other. A similar two-lobe structure has been found in the transient dynamics of the Wigner function for the dissipative quantum Rabi model with an additional Stark shift term included\cite{algr13}. For the closed system, this behavior readily follows from a sort of mean-field approach identical to the Born-Oppenheimer approximation~\cite{jlar07,boa}. In particular, within this approximation and for $\gamma=0$ the maxima of the Wigner function are found for
\begin{equation}\label{dickecrit}
\mathrm{Im}(\alpha)=\left\{
\begin{array}{lll}
0,& \hspace{0.5cm} & g\leq g_c\\ \\
\pm\sqrt{\frac{g^2}{\omega^2}-\frac{\Omega^2}{16g^2}}, & \hspace{0.5cm} & g>g_c
\end{array}\right.,
\end{equation}
where $g=(\lambda_1+\lambda_2)/2$ and $g_c=\sqrt{\omega\Omega}/2$, and $\mathrm{Re}(\alpha)\equiv0$. Note that $g_c$ coincide with the critical coupling of the Dicke model~\cite{hepp}. Thus, the splitting grows approximately linear with $\lambda_1+\lambda_2$. If the signs of the couplings $\lambda_1$ and $\lambda_2$ are different the roles of the real and imaginary parts of $\alpha$ are interchanged. Now, Fig.~\ref{Wigner_all_ratios} shows that the real parts for the locations of the Wigner function maxima are actually non-zero. This is seen in the anti-clockwise tilting of the Wigner function. This is a result of the Lamb shift~\cite{smba97}, i.e. it only arises due to the coupling to the bath. This is also easily understood by considering the mean-field versions of the Heisenberg equations of motion for the open Rabi model. The steady state solution for the field will then include a non-zero imaginary part (or real part depending on how one defines $\alpha$) which implies that the field phase for the separated blobs is not exactly $\pm\pi$~\cite{morigi}.

\section{Model B - Anisotropic Dicke model (normal phase)}
\label{sec:model2}
The Rabi model is highly quantum in the sense that quantum fluctuations play a dominant role in the spin sub-space, and consequently it lacks a natural classical limit. Such a limit would, however, emerge if we allow the spin to become very large, i.e. the Pauli matrices are replaced by general angular momentum operators $\hat S_\alpha$ ($\alpha=x,\,y,\,z$). In doing this replacement we obtain the anisotropic Dicke model (ADM)
\begin{equation}\label{dicke}
\hat H_{\rm ADM}\!=\! \omega \hat{a}^{\dagger}\hat{a}+\!\frac{\Omega}{2} \hat{S}_{z} +\lambda_1\!\left(\hat{a}^{\dagger}\hat{S}^{-}\!+\!\hat{S}^{+}\hat{a}\right)+\!\lambda_2\!\left(\hat{a}^{\dagger}\hat{S}^{+}\!+\!\hat{S}^{-}\hat{a}\right)\!.
\end{equation}
It is the purpose of this section to demonstrate that non-classicality of the steady state survives also in this classical limit.

With the correct scaling of the coupling parameters (relative to the system size), the Dicke model is critical with the normal phase characterized by a vacuum photon field and the spin pointing towards the south pole. The superradiant phase instead comprises a coherent photon state and the spin rotated away from the south pole~\cite{hepp}. True criticality only emerges in the thermodynamic limit meaning that the spin $S\rightarrow\infty$~\cite{dickefin}. In the strict limit of infinite spin, the spectrum is linear and we may replace spin operators by boson operators (plus an overall energy shift). In such a case we consider
 the following Hamiltonian 
\begin{equation}\label{lincpledosc}
\hat H_{\rm nAD}=\omega \hat{a}^{\dagger}\hat{a}+\Omega \hat{b}^{\dagger}\hat{b} +\lambda_1\left(\hat{a}^{\dagger}\hat{b}+\hat{b}^{\dagger}\hat{a}\right)+\lambda_2\left(\hat{a}^{\dagger}\hat{b}^{\dagger}+\hat{b}\hat{a}\right),
\end{equation}
where the notation $\hat H_{\rm nAD}$ refers to the linearized anisotropic Dicke model in the normal phase. Clearly the above Hamiltonian can be used to model various other physical systems, including coupled harmonic oscillators/cavity modes \cite{cjos14}. Going from the non-linear model (\ref{dicke}) to the linear approximated model (\ref{lincpledosc}) allows for an analytical approach. For finite systems the non-linearity will enter after sufficiently long times~\cite{bye}. Here, however, we assume that this time-scale is much longer than the intrinsic relaxation time of the open system, i.e. we push any quantum revivals to happen beyond our experimental times. In the normal phase of the Dicke model, the linear Hamiltonian $\hat H_\mathrm{nAD}$ describes the collective quantum excitations/fluctuations (Bogoliubov modes)~\cite{fdim07}. This is most easily seen by linearizing the original Dicke model in the Holstein-Primakoff boson representation~\cite{fdim07,dickefin}. In the superradiant phase, the resulting quadratic boson Hamiltonian contains additional squeezing terms, e.g. $\hat b^2$ and $\hat a^2$ \cite{cjos15}. Within the effective linear model, the quantum fluctuations cause entanglement in the system, and for the closed Dicke model it has been found that the atom-field entanglement peaks at the critical point~\cite{dickeent}. We point out, however, that since the Lindblad jump operator $[\hat a,\hat H_\mathrm{ADM}]\neq0$ it is not, in principle, clear that criticality of $\hat\rho_\mathrm{ss}$ is preserved in the presence of the photon decay. This issue was recently addressed and photon losses do not destroy criticality but alters both the critical coupling and the critical exponents~\cite{fdim07,opendicke}. 
As discussed in more details below, the conservation of the critical point under photon losses can be understood from the fact that the state for the normal phase is actually a dark state of the Lindblad operators. Here, as we are interested in the opposite (classical) limit of the Rabi model, we only consider the fluctuations in the normal phase and not in the superradiant one. This emerges in the present analysis as destabilization of the Bogoliubov modes upon approaching the critical point.

 \begin{figure}[h!]
  \centering
    \includegraphics[width=0.45\textwidth]{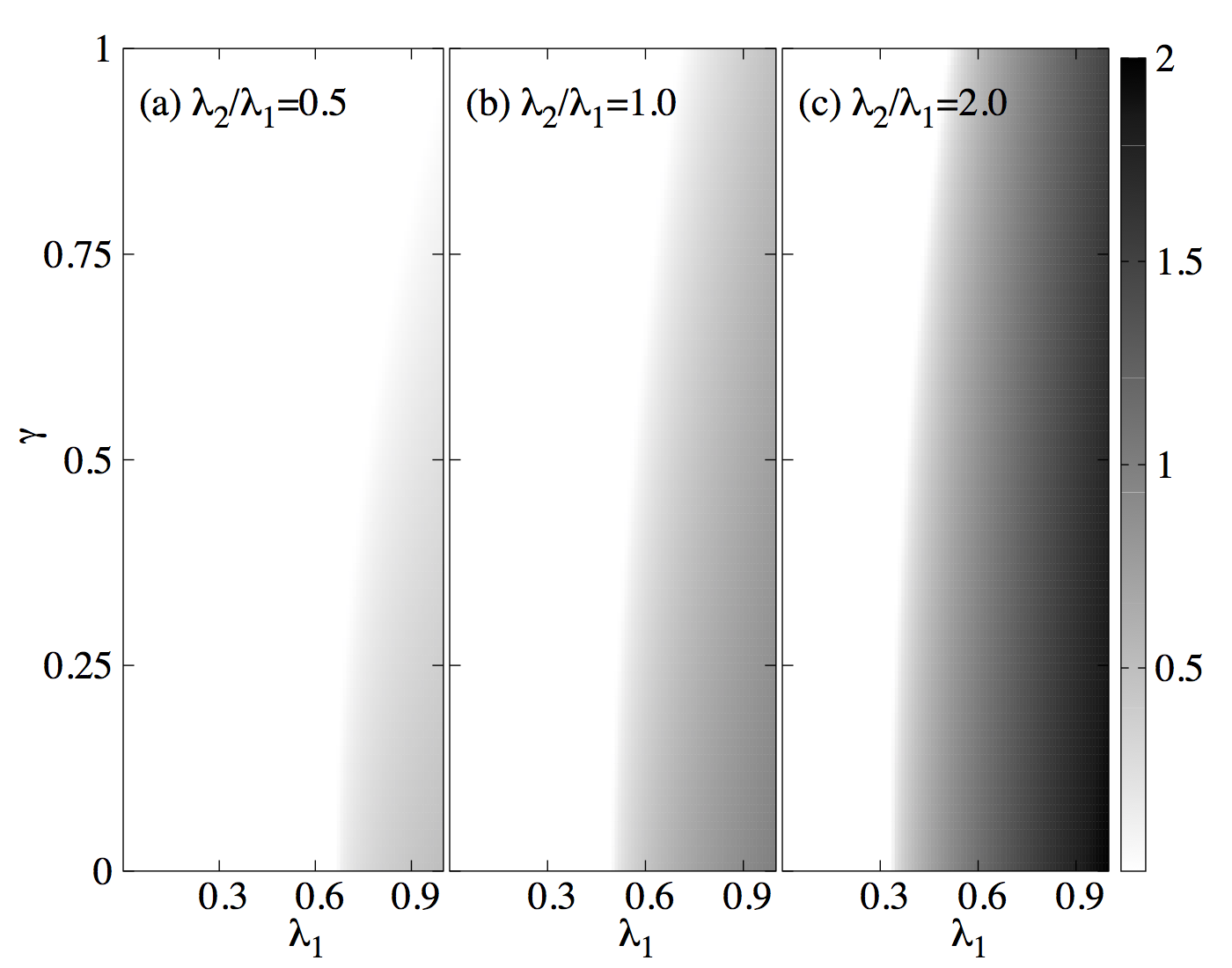}
    \caption{Stability criterion $\zeta$ \eqref{stbparm} for different values of the ratio $\lambda_2/\lambda_1$,  shown here as a function of the decay rate of the field mode $\gamma$ and the coupling strength $\lambda_1$. The non-zero value $\zeta\neq0$ mark the transition to the unstable regime. The dimensionless atomic and field frequencies $\Omega=\omega=1$.}\label{stblinosc}
\end{figure}

Going beyond the study in~\cite{dickeent}, we here include the interaction with the outside environment. We again envisage a scenario where damping of the field  mode is the only dominant dissipation channel;
\begin{equation}\label{openlincpledosc}
\dot{\hat{\rho}}= -i[\hat H_{\rm nAD},\hat\rho]+\gamma\mathcal L_{\hat{a}}\hat{\rho}.
 \end{equation}
In the above master equation we have neglected dissipation of the mode $\hat{b}$. We again argue that this assumption can be justified  if it is possible to realize the ADM  \eqref{lincpledosc}~through a  scheme outlined in \cite{fdim07}. Specifically, mode $\hat{b}$ can play the role of collective Bogoliubov excitations of a collection of ``two-level'' systems, where each ``two-level'' system can be the metastable  low lying energy doublet of a Raman driven four level atom.  Also, as discussed in the previous section, a phenomenological introduction of the Lindblad super operator for the field mode as introduced in the  master equation \eqref{openlincpledosc} can  only be justified for an original  time-dependent Hamiltonian. It is worth pointing out that  for a time-independent model described by the Hamiltonian  \eqref{lincpledosc} a correct description of the open dynamics results in a non-local master equation which disagrees with the predictions of the master equation \eqref{openlincpledosc} \cite{cjos14}.

\begin{figure}[h!]
  \centering
    \includegraphics[width=0.45\textwidth]{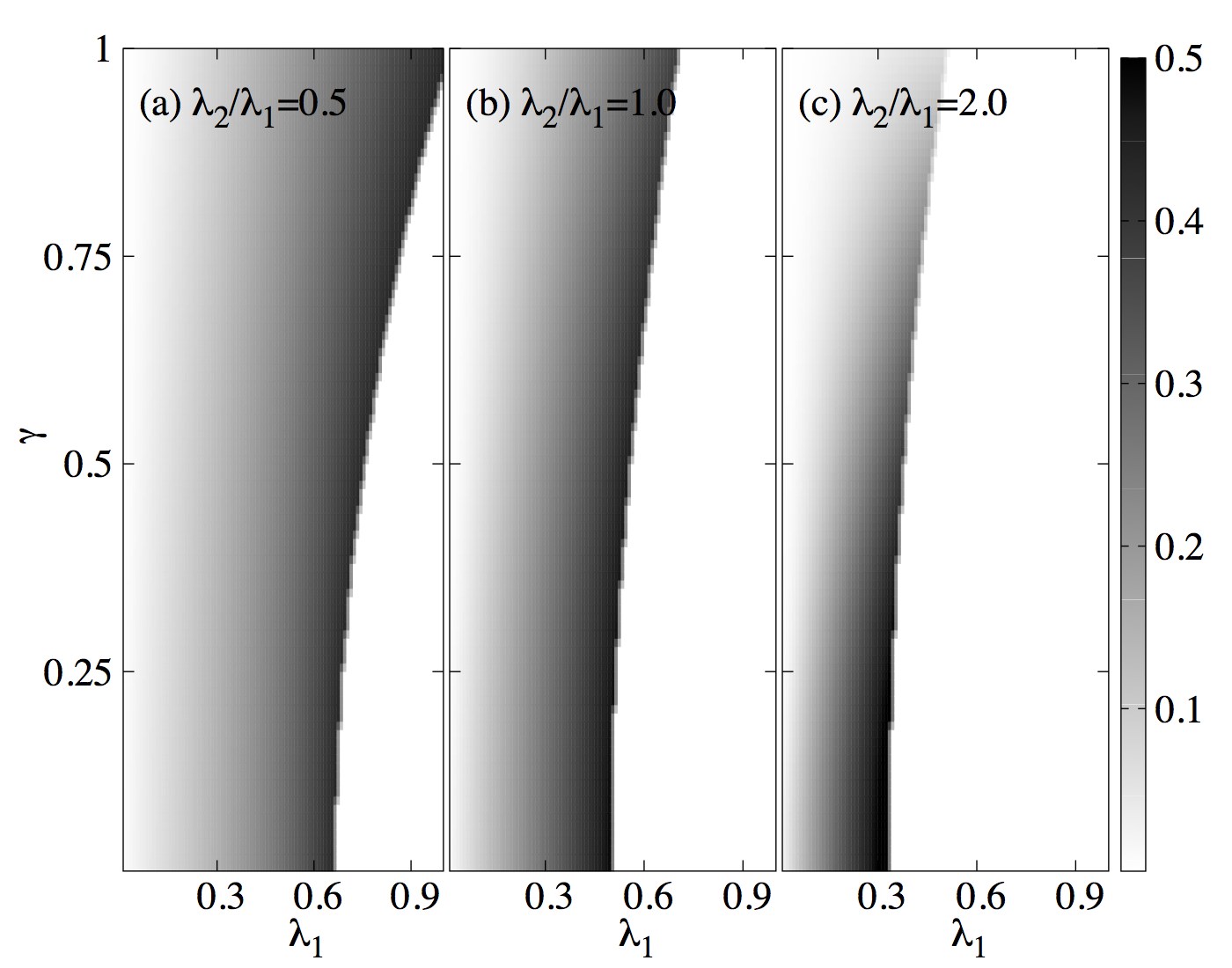}
    \caption{Bi-partite entanglement between the modes $\hat{a}$ and $\hat{b}$ for three different values $\lambda_2/\lambda_1$ and as a function of $\gamma$ and $\lambda_1$. Right before the solution gets unstable (white region, and see previous figure \ref{stblinosc}) the entanglement is maximized. This is in agreement with the fact that in the full Dicke model this marks a critical point~\cite{dickeent}. Note, however, that the critical point has been shifted due to the coupling to the bath. The remaining dimensionless parameters are $\Omega=\omega=1$.}\label{entlinosc}
\end{figure}

As already mentioned, since the Hamiltonian is quadratic we may approach the problem analytically. To this end we define the two-mode normal-ordered characteristic function $\chi(\epsilon_{a},\epsilon_{b},t)=\langle e^{\epsilon_{a}\hat{a}^{\dagger}}e^{-\epsilon_{a}^{*}\hat{a}}e^{\epsilon_{b}\hat{b}^{\dagger}}e^{-\epsilon_{b}^{*}\hat{b}}\rangle$. From  the quantum characteristic function the complete statistical description of the corresponding state can be obtained. One can therefore obtain the expectation values of quantum mechanical observables, e.g. for the equal time correlator one has
\begin{multline*}
	\langle \hat{a}^{\dagger m}(t){\it \hat{b}^{\dagger n}}(t) \rangle\\=\it {\left(\frac{\partial }{\partial \epsilon_{a}}\right)^{m}\left(\frac{\partial }{\partial \epsilon_{b}}\right)^{n}\chi(\epsilon_{a},\epsilon_{a}^{*},\epsilon_{b},\epsilon_{b}^{*},{\it t})}|_{\epsilon_{a},\epsilon_{a}^{*},\epsilon_{b},\epsilon_{b}^{*}=\rm{0}}.
\end{multline*} 
For an initial Gaussian state of the two modes $\hat{a}$ and $\hat{b}$, the master equation \eqref{openlincpledosc} can be expressed  as  a partial differential equation for  $\chi$ \cite{cjos14} 
 \begin{equation}\label{chieqn}
\frac{\partial }{\partial t} \chi
=z^{T}  {\mbox{\bf M}} z \chi+z^{T}  {\mbox{\bf N}} \nabla \chi,
\end{equation}
where $z^{T}=( \epsilon_{a},\epsilon_{a}^{*},\epsilon_{b},\epsilon_{b}^{*}), \nabla= (\frac{\partial}{ \partial \epsilon_{a}}, \frac{\partial }{ \partial \epsilon_{a}^{*}}, \frac{\partial}{ \partial \epsilon_{b}}, \frac{\partial }{ \partial \epsilon_{b}^{*}})^{T}$
and
\begin{eqnarray*}
{\mbox{\bf M}}&=&\left( \begin{array}{cccc}
  0 & 0 & i \lambda_{2}/2&0 \\
     0 & 0 & 0&-i \lambda_{2}/2\\
      i \lambda_{2}/2 & 0 & 0&0 \\
       0 & -i \lambda_{2}/2 &0 &0 \\
  \end{array} \right ),\\ \\
{\mbox{\bf N}}&=&\left( \begin{array}{cccc}
   i\omega-\gamma & 0 & i \lambda_{1} &-i\lambda_{2} \\
     0 & -i\omega-\gamma& i \lambda_{2} &-i \lambda_{1} \\
      i\lambda_{1}  & -i\lambda_{2} & i\Omega&0 \\
       i\lambda_{2}  & -i \lambda_{1}  &0 &-i\Omega\\
  \end{array} \right )     
  \end{eqnarray*}
  are the drift and diffusion matrices respectively. We first analyze the steady state solution of the master equation~\eqref{openlincpledosc}. In particular, the solution is stable if the real parts of all the eigenvalues of the diffusion matrix ${\mbox{\bf N}}$ are all negative. In particular, when $\Omega=\omega$, the stability of the steady state requires
  \begin{equation}\label{stbparm}
  \zeta={\rm Max ~\Re({\pm}\nu^{+},\pm\nu^{-}}) -\gamma<0,
  \end{equation}
  where, 
   \begin{eqnarray}
  \nu^{+}&=& \sqrt{\gamma ^2-4 \left(\lambda_{1} ^2-\lambda_{2}^2+\omega ^2+\sqrt{4 \lambda_{1} ^2 \omega ^2-\gamma ^2 \omega ^2} \right)}, \nonumber \\
 \nu^{-}&=&  \sqrt{\gamma ^2-4 \left(\lambda_{1} ^2-\lambda_{2}^2+\omega ^2-\sqrt{4 \lambda_{1} ^2 \omega ^2-\gamma ^2 \omega ^2} \right)}. \nonumber 
  \end{eqnarray}
  
The stability parameter $\zeta$ is shown in Fig.~\ref{stblinosc} as a function of the damping rate of the field mode $\gamma$ and the two-mode  coupling strength $\lambda_1$. As is evident from the figure, increasing the damping of the field mode allows a stable  steady state to be achieved for an extended region of parameter space spanned by $\lambda_1$. This behavior can be understood physically by considering the different terms of the master equation \eqref{openlincpledosc}. First we may note that the instability of the mode coincides with the Dicke phase transition. Indeed, the model Hamiltonian (\ref{lincpledosc}) describes the collective excitations in the normal phase and not in the superradiant phase, and hence, these excitation modes cannot be analytically continued into the superradiant phase. Thus, we conclude that increasing the photon loss  gives a more extended normal phase. This is expected since the Lindblad term of Eq. (\ref{openlincpledosc}) favors the vacuum state of the field, which is also  favored by the bare field energy $\omega\hat a^\dagger\hat a$. On the other hand, the atom-field coupling term is not minimized by such a field state. Without photon decay, the transition occurs at $\left(\lambda_1+\lambda_2\right)=\sqrt{\omega\Omega}$ (see Eq.~(\ref{openlincpledosc})), but now losses support the normal phase and thereby shift the critical value. Following Refs.~\cite{fdim07,dickeopen} one explicitly finds $\lambda_c=\sqrt{\Omega\left(\omega^2+\gamma^2\right)/\omega}/2$ which is obtained at the mean-field level, i.e. in the thermodynamic limit. In the Dicke limit ($\lambda_1=\lambda_2=\lambda$) it is easy to obtain the phase boundary $\omega^{2}+\gamma^{2}=\mathrm{4}\lambda^{2}$, which retrieves the result for the critical coupling of the open Dicke model on resonance \cite{fdim07,dickeopen}.

The steady state of  light-matter coupled quantum system modeled under the master equation \eqref{rabianisoopen} was shown to be an inseparable state of the cavity field and the two-level system. Somewhat surprisingly, this observation can also be extended to the steady state of the normal phase of the Dicke model \eqref{openlincpledosc}. In particular, one could argue that in the thermodynamic limit as studied here any entanglement should vanish since for the Dicke model quantum fluctuations are negligible~\cite{sachdev}. That this is not the case derives from the presence of the critical point. To demonstrate sustainable quantum correlations we next evaluate the  steady state bi-partite entanglement between the two modes $\hat{a}$ and $\hat{b}$ and characterize it in terms of the logarithmic negativity which for two-mode Gaussian states serves as a necessary and sufficient criterion for the inseparability~\cite{gera07}. A two-mode Gaussian state can be fully quantified in terms of its covariance matrix ${\mbox{\bf V}}$ which is a $4 \times 4$ symmetric matrix with $V_{ij}=(\langle R_{i} R_{j}+R_{j} R_{i} \rangle )/2$ and $R^{T} =(\hat{q}_{a}, \hat{p}_{a},\hat{q}_{b}, \hat{p}_{b})$. Here $\hat{q}_{a,b}$ and $\hat{p}_{a,b}$ are the position and momentum quadratures of the mode $\hat{a}(\hat{b})$. For a two-mode Gaussian continuous-variable state with covariance matrix $\bf V$, the logarithmic negativity is obtained as $ \mathcal N= \rm{Max}[0,-\rm{log}(2 \nu_{-})]$ \cite{gera07}, where $\nu_{-}$ is the smallest of the symplectic eigenvalues of the covariance matrix, given by $\nu_{-}= \sqrt{\sigma/2-\sqrt{(\sigma^{2}-4 \rm{Det} \bf{V})}/\rm{2}}$. Here
\begin{eqnarray*}
\sigma&=&\rm{Det} \bf{A_{1}}+\rm{Det} \bf{B_{1}}-\rm{2Det} \bf{C_{1}}\\ \bf V&=&
\left(
\begin{array}{cc}
  \bf A_{1} &\bf C_{1}\\
 \bf C_{1}^{T} & \bf B_{1}
\end{array}
\right),
\end{eqnarray*} 
where $\bf {A_{1}}$ $(\bf {B_{1}})$ accounts for the local variances of mode $a$ $(b)$ and $\bf C_{1} $ for the inter-mode correlations. Using the numerical solutions of the partial differential equations \eqref{chieqn} we compute the logarithmic negativity. The steady state two-mode entanglement is shown in Fig.~\ref{entlinosc} where it is given as a function of the damping rate $\gamma$  and the two-mode  coupling strength $\lambda_1$. As is evident, the maximum attainable bi-partite entanglement between the two modes monotonically increases on increasing the value of the ratio $\lambda_2/ \lambda_1$. The highest entanglement is obtained at the critical point, which is also known from the closed Dicke model~\cite{dickeent}. This is, in fact, a general property for quantum phase transitions; at the critical point where characteristic length scales diverge, the entanglement also diverges~\cite{ptent}. However, already for the closed Dicke model the phase transition is not a `typical' quantum phase transition since there is no length scale in the problem and, as we mentioned above, quantum fluctuations vanish in the thermodynamic limit. Because of this, the transition has been called `classical'~\cite{sachdev}. As such, the results of Fig.~\ref{entlinosc} suggest that, qualitatively, the same behavior of the entanglement is found for a dynamical phase transition of an open driven system.

\section{Quantum control in hybrid architectures}
\label{sec:feedbck}

 \begin{figure}[h!]
  \centering
    \includegraphics[width=0.5\textwidth]{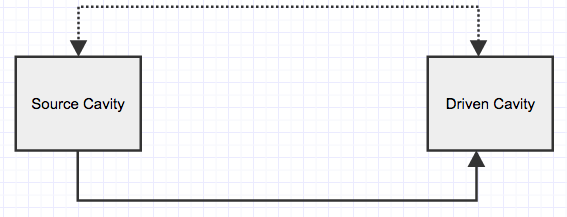}
    \caption{Schematic plot of the feedback loop between the two cavities.  We envision a scenario of two coupled cavities;  a source cavity and a driven cavity. The driven cavity is assumed to be slaved to the source cavity. The source cavity drives the state of the slave cavity by means of a unidirectional coupling (shown by a solid line). The driven cavity in turn also influences the state of the source cavity by means of a reversible interaction (shown by a dotted line). The simultaneous presence of reversible and irreversible couplings between the cavity modes results in an all-optical feedback loop.}\label{setupfedbck}
\end{figure}

 \begin{figure}[h!]
  \centering
    \includegraphics[width=0.5\textwidth]{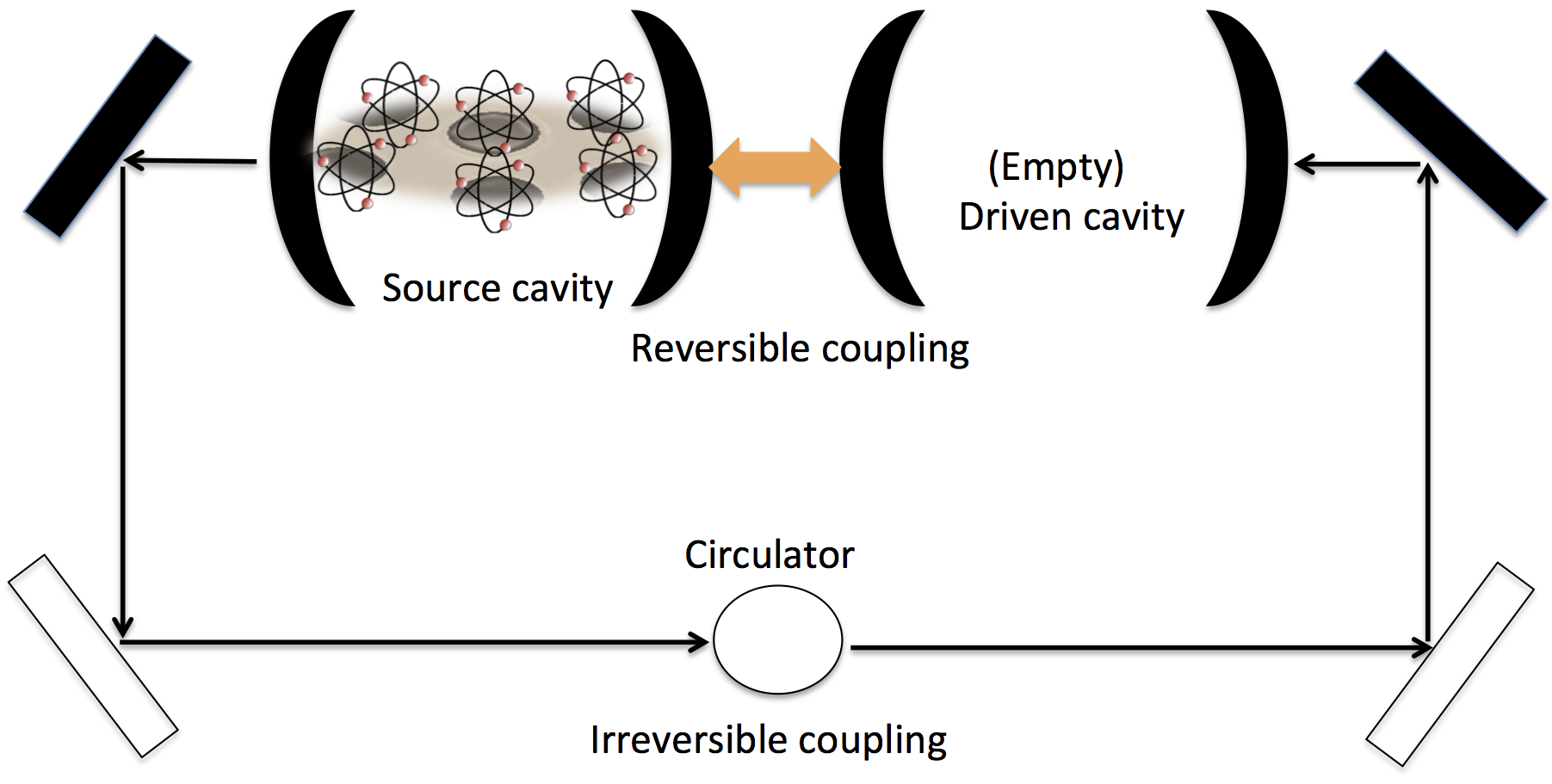}
    \caption{Illustration of the general all-optical feedback of Fig.~\ref{setupfedbck} specifically used for establishing quantum control in the Dicke model of Section \ref{sec:model2}. The internal dynamics of the source cavity is modeled  by \eqref{lincpledosc} while the driven cavity is initially prepared in its vacuum state. Mode $\hat{a}$ of the source cavity and the mode $\hat{c}$ of the driven cavity are interacting under a reversible interaction  of the form \eqref{revint}. Mode $\hat{a}$ of the source cavity  also couples irreversibly to mode $\hat{c}$ of the driven cavity. Such a {\it feed-forward} coupling can be engineered through feeding the mode $\hat{a}$ from one end of the source cavity  and reflecting it onto the driven cavity through a series of mirrors (filled) and beam-splitters (unfilled). Optical circulators or Faraday isolators can be used  to prevent interference from reflections in the opposite directions. Reversible and irreversible interactions jointly constitute an all-optical feedback. If this feedback loop has negligible time delay a Markovian master equation to describe the open dynamics of the source cavity can be derived \eqref{finlmseqn} as detailed in Section \ref{sec:feedbcknodely}.}\label{speccficscheme}
\end{figure}

In the previous two sections we have studied the open dynamics of the anisotropic Rabi and Dicke models and have explicitly explored the properties of  the NESSs of these models. However, so far we have eluded ourselves from effectively engineering the open dynamics of these hybrid models. Even though it was found that entanglement persists in both models despite losses, in the present section we wish to explore the possibilities to enhance the amount of entanglement by actively introducing feedback into the model.
 
An unusual ingredient which enriches the dynamics of quantum systems is their inherent  `openness' which may result in exotic NESS properties 
\cite{fver09,jtba11,ylin13,cjosh13,msch15}. This was in particular the theme of the previous two sections. However, engineering a controlled degree of dissipation is also an important   tool in realizing tasks of quantum information processing. In this spirit  we will now use an all-optical feedback scheme for quantum state protection and  establishing quantum control in the normal phase of the Dicke model. The scheme we choose is based on coherent feedback, i.e. no measurement is performed in the feedback loop \cite{hmwi10}.  A symbolic plot of a coherent (all-optical) feedback scheme is illustrated in Fig.~\ref{setupfedbck}; it comprises of two units, which have been labeled source and driven cavities, with reversible and irreversible couplings between them. The simplest all optical feedback loop will use the  output from a  source cavity and feedforward  into a driven cavity which is coupled to the source cavity in some way. It is worth mentioning that unidirectional and bidirectional couplings between the source and the driven cavities jointly constitute an all optical feedback scheme. A coherent feedback loop, however, can also introduce a finite time-delay $\tau$.  Firstly, assuming that the time-delay $\tau$ introduced by the feedback loop is negligible, we study the open dynamics of the source cavity \cite{hmwi94,ptom01,cjosnj14}. Subsequently, making use of a time-delayed feedback control method, \cite{kpyrg92,nyam14,algr14} we incorporate a time-delay $\tau$ introduced by  the feedback loop.

\subsection{All optical feedback loop with negligible time-delay $\tau$}
\label{sec:feedbcknodely}
We now apply the general scheme of all-optical feedback illustrated in Fig.~\ref{setupfedbck} for a specific task of quantum state protection in  the anisotropic  Dicke model of Sec. \ref{sec:model2}. A schematic of our proposal to implement all-optical feedback is shown in Fig.~\ref{speccficscheme}. The internal (unitary) dynamics of the source cavity is described by the Hamiltonian \eqref{lincpledosc}, i.e. $\hat H_{\rm source}=\hat H_{\rm nAD}$ and the driven cavity is reversibly coupled to the field mode of the source cavity by the following Hamiltonian \cite{hmwi94,ptom01}
\begin{equation}\label{revint}
\hat H_{\rm int}=i\mu\frac{\sqrt{\gamma\gamma_\mathrm{d}}}{2}(\hat{a}^{\dagger}\hat{c}-\hat{c}^{\dagger}\hat{a}),
\end{equation}
where $\mu$ is a  dimensionless coupling parameter, $\hat{c}^{\dagger}$ and $\hat{c}$ are  the creation and annihilation operators for the field mode of the driven cavity which has damping rate $\gamma_\mathrm{d}$. The driven cavity is assumed to be prepared in its vacuum state with its respective internal dynamics modeled as $\hat H_\mathrm{driven}=\Omega_\mathrm{d}\hat{c}^{\dagger}\hat{c}$. A reversible interaction  of the form \eqref{revint} can arise through mode overlap between the modes $\hat{a}$ and $\hat{c}$~\cite{hartmann}.

Under the Born-Markov approximation a joint state of the source and driven cavities, represented here as $\hat{W}$, evolves under the following master equation \cite{hmwi10,hmwi94,ptom01}
\begin{eqnarray}
\dot{\hat{W}}&=&-i\left[\hat H_{\rm nAD}+H_{\rm driven}+\hat H_{\rm int},\hat{W}\right]\nonumber \\
&&+\sqrt{\gamma\gamma_\mathrm{d}}([\hat{a}\hat{W},\hat{c}^{\dagger}]+[\hat{c},\hat{W}a^{\dagger}])\nonumber \\
&&+\frac{\gamma}{2}\mathcal L_{\hat{a}}\hat{W}+\frac{\gamma_\mathrm{d}}{2} \mathcal L_{\hat{c}}\hat{W}.
\end{eqnarray}
It should be remarked that $\hat{W}$ represents a  tri-partite state of modes $\hat{a},\hat{b}$, and $\hat{c}$. On the grounds of arguments presented previously, we have again neglected damping of the  mode $\hat{b}$ of the source cavity. The two terms appearing in the second line account for the unidirectional coupling between the source and  driven cavities and the last two terms are the individual Lindblad operators describing photon losses for the source and driven cavity modes respectively. 
As  illustrated in Fig.~\ref{speccficscheme}, such a unidirectional coupling between the source and driven cavities can be established using an optical circulator \cite{hmwi94}: 
a non-reciprocal optical device such as Faraday rotator can be used to establish irreversible coupling between the optical modes $\hat{a}$ and $\hat{c}$ of the source and the driven cavities. As pointed out in Refs.~\cite{hmwi94,ptom01}, for  the feedback loop to be effective the driven cavity should respond much faster than the source cavity. We thus work in a regime where $\gamma_\mathrm{d}\gg1$  meaning that the state of the driven cavity is slaved to the state of the source cavity. Since $\gamma_\mathrm{d}$ is the large parameter (i.e. determining the fast time scale) and the driven cavity is assumed to couple to a zero temperature reservoir, the population of the driven cavity mode can be truncated to only the lowest photon states. We therefore approximate the joint state of the two coupled cavities as 
\begin{eqnarray}\label{appstate}
\hat{W}&=&\hat{\rho}_{00}|0\rangle \langle 0|+\hat{\rho}_{10}|1\rangle \langle 0|+\hat{\rho}_{10}^{\dagger}|0\rangle \langle 1|\nonumber \\
&& +\hat{\rho}_{11}|1\rangle \langle 1|+\hat{\rho}_{20}|2\rangle \langle 0|+\hat{\rho}_{20}^{\dagger}|0\rangle \langle 2|,
\end{eqnarray}
where $\hat{\rho}_{ij}$ is the conditional state of the source cavity (a joint state of modes $\hat{a}$ and $\hat{b}$) when the driven cavity is projected on the state space $|i\rangle \langle j|$. Using the above ansatz and adiabatically eliminating the driven cavity it is possible to derive an effective master equation for the source cavity alone \cite{hmwi94,ptom01}. Following the derivation provided in Appendix \ref{appendix1} one arrives at the resulting master equation 
\begin{eqnarray}\label{finlmseqn}
\dot{\hat{\rho}}&=&\dot{\hat{\rho}}_{00}+\dot{\hat{\rho}}_{11}=
 -i\left[\hat H_{\rm nAD},\hat{\rho}\right]+\frac{\gamma_{\rm eff}}{2}\mathcal L_{\hat{a}}\hat{\rho},
 \end{eqnarray}
 where $\gamma_{\rm eff}=\gamma\left(1+\mu(2+\mu)\right)$ is the new effective damping rate of the field mode. On choosing the dimensionless parameter $\mu=-1$, it is remarkable to observe that  an all optical feedback loop is capable of completely blocking the loss of the source cavity. However, losses in the feedback loop would deteriorate the effectiveness of such feedback. If the feedback loop has an efficiency  $\eta$ ($\le 1$),  then it is possible to show that the  effective damping rate of the source cavity gets modified as  $\gamma_{\rm eff}=\gamma \left(1+\eta \mu(2+\mu)\right)$ \cite{ptom01}. We therefore conclude that,  under the assumptions that the time-delay introduced by a feedback loop is negligible and  the  state of the driven cavity is effectively slaved to the source cavity,  it is possible to establish arbitrary control over the damping rate of the source cavity. This can be an important step for quantum state protection in hybrid quantum systems. 
 
\begin{figure}[h!]
  \centering
    \includegraphics[width=0.48\textwidth]{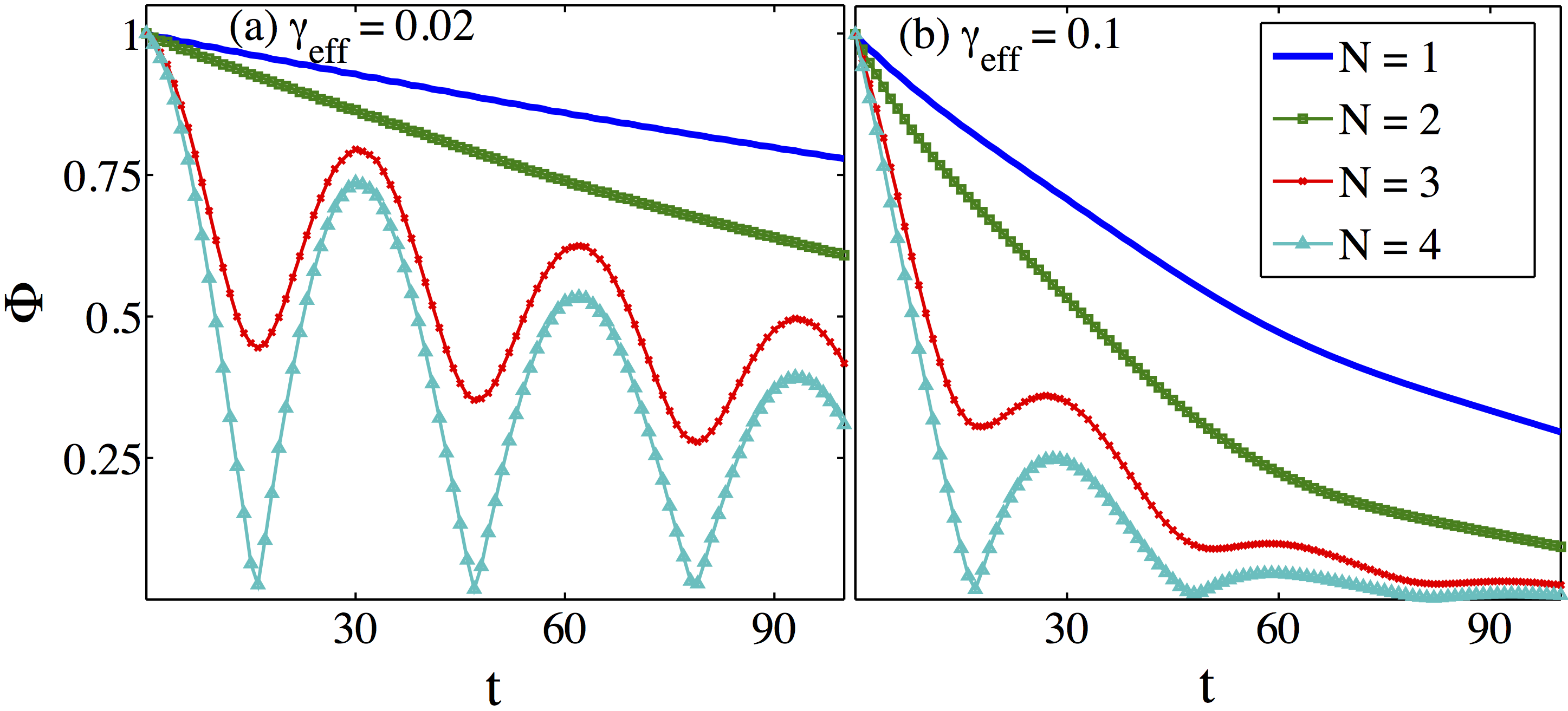}\\
     \includegraphics[width=0.48\textwidth] {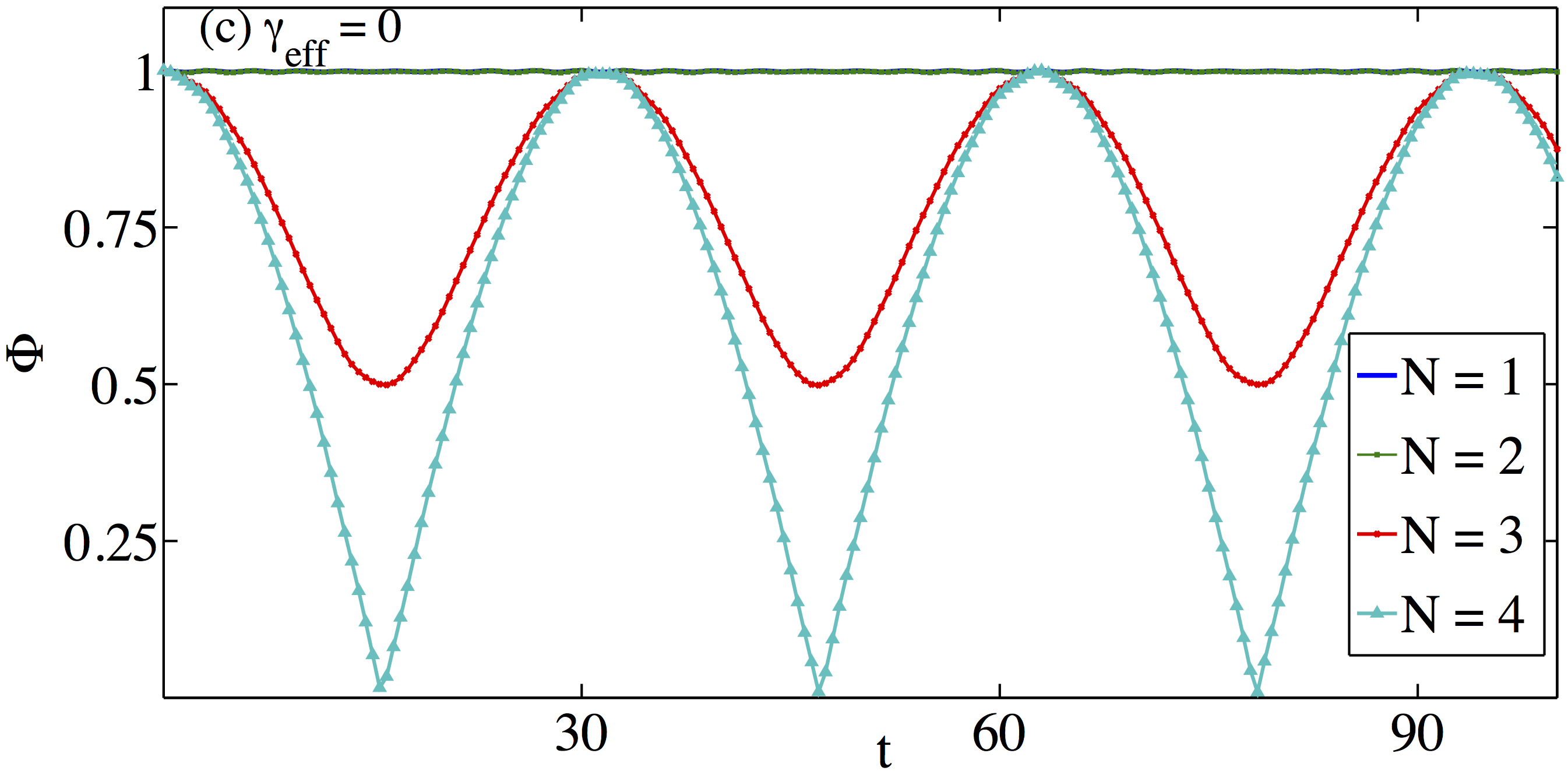}\\
     \includegraphics[width=0.48\textwidth] {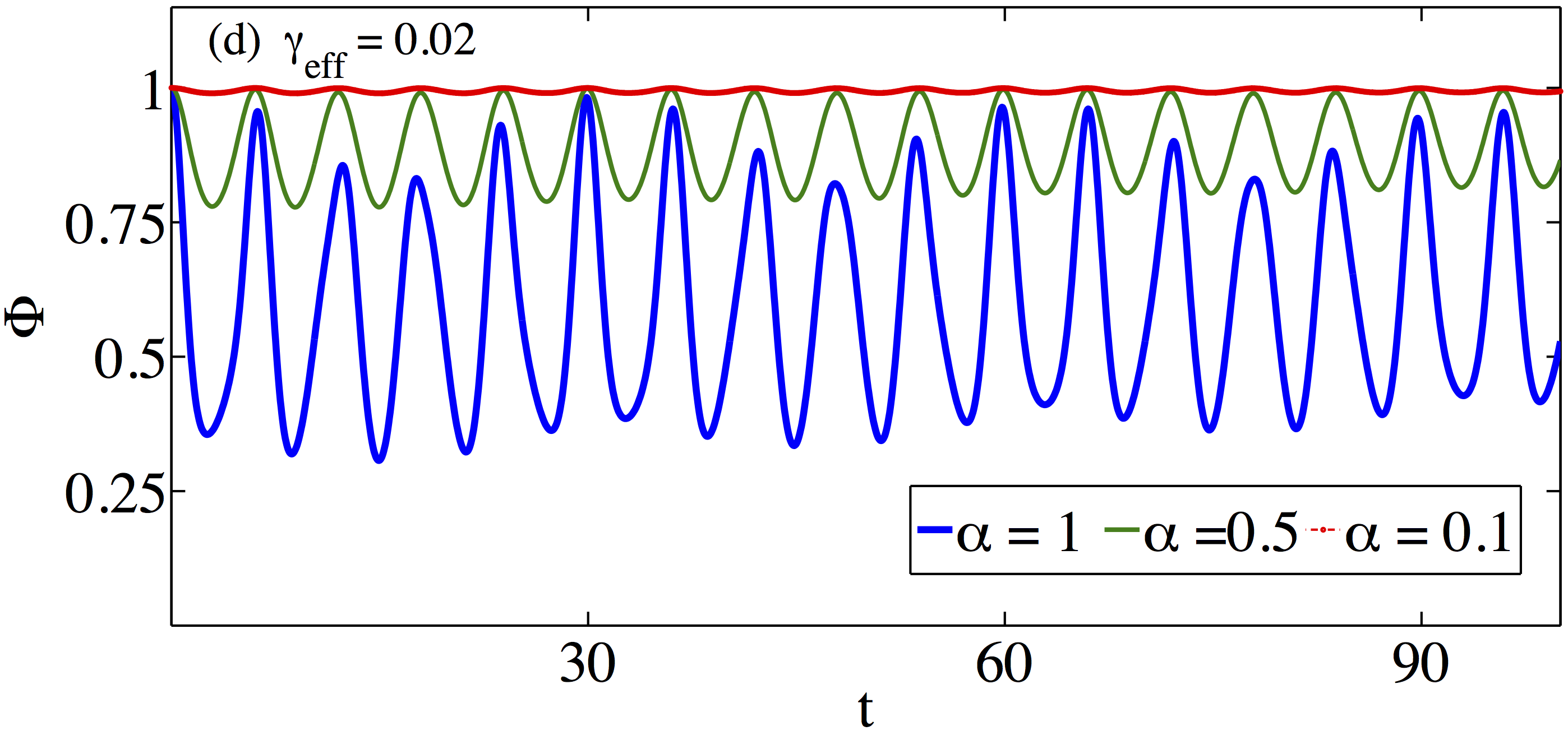}
        \caption{(Color online) The upper plots (a) and (b) give the time evolution of the fidelity $\Phi$ for  a bi-modal NOON state in the presence of losses, while (c) shows the same but in absence of photon losses. By comparing the first two examples it is clear that the feedback loop makes it possible to preserve coherent evolution for longer times. The last plot (d) is the same but for an entangled cat state with one mode in vacuum and the other in a coherent state $|\alpha\rangle$. The shorter period results from the dephasing of the different Fock states involved. The other dimensionless parameters are $\Omega=\omega=1, \lambda_1=0.05$ and $\lambda_2/\lambda_1=1$ in all four panels.}
        \label{fidtimevolution}
\end{figure}

As an example application of an all optical feedback scheme for quantum state protection, we assume that the two coupled modes interacting under the Hamiltonian \eqref{lincpledosc} are initially prepared in a NOON state~\cite{noon}. NOON states are Bell-like entangled states with applications in quantum metrology and quantum lithography and they may also be utilized for achieving phase supersensitivity \cite{vgio04}. We assume that the modes $\hat{a}$ and $\hat{b}$ are initialized in a state  $|\Psi_{-}^{N} \rangle =(|N\rangle_{a} |0 \rangle_{b}-|0 \rangle_{a}N \rangle_{b} )/\sqrt{2}$.  We examine the autocorrelation function, i.e. the overlap between $\hat\rho(0)$ and the time evolving joint state of the two modes  $\hat\rho(t)$, which for mixed states is given by the Uhlmann fidelity \cite{auhl76,rjoz94} 
\begin{equation}\label{uhmnfid}
\Phi=\rm{Tr} \sqrt{\sqrt{\hat\rho(0)}\hat\rho(t)\sqrt{\hat\rho(0)}}.
\end{equation}
The time evolution of the fidelity $\Phi$ for different values of $N$ and two  different values of $\gamma_{\rm eff}$ is plotted in Fig.~\ref{fidtimevolution} (a) and (b). We note that the results are for a small coupling ($20\lambda_1=20\lambda_2=\omega=\Omega$) in order to assure numerical convergence. In Fig.~\ref{fidtimevolution} (c) we also show the time evolution of the fidelity $\Phi$ for an initial bi-model NOON state in absence of photon losses. As is evident, by controlling the damping rate of the source field it is possible to prolong the lifetime of the initially prepared NOON state. It follows that suitably choosing $\gamma_{\rm eff}$ allows the value of the fidelity to be kept above its classical value of 2/3 \cite{smas95} for a longer time. It is also worth noting that  NOON states become more susceptible to the environmental damping with the  increase in the excitation number $N$. In this scenario establishing a control over the damping rate of the field in the source cavity could be an important step in preserving initial quantum coherence. In the Appendix~\ref{appendix2} we analytically analyze the structure of especially plot (c), and explain why the fidelity stays constant for $N=1,\,2$ and not for $N=3,\,4$.

Also shown in Fig.~\ref{fidtimevolution} (d) is the evolution of the Uhlmann fidelity of a two-mode entangled coherent state $|\Psi \rangle =(|\alpha\rangle_{a} |0 \rangle_{b}+|0 \rangle_{a}|\alpha \rangle_{b} )/\sqrt{2(1+e^{-|\alpha|^{2}})}$, where $|\alpha\rangle$ represents a coherent state. Entangled coherent states can be a useful resource in quantum metrology and can exhibit  noticeable improved sensitivity for phase estimation when compared to that for NOON states. Entangled coherent states can also  outperform the phase enhancement achieved by NOON states both in the lossless, weak, moderate and high loss regimes \cite{jjoo11}.  As demonstrated in Fig.~\ref{fidtimevolution} (d), when compared to a NOON state a two-mode entangled coherent state is found to be more resilient to photon losses. Nevertheless, the quantum fidelity of entangled coherent states also seems to  decay significantly  with the increase in the mean number of photons $|\alpha|^{2}$. 

\subsection{All optical feedback loop with finite time-delay $\tau$}
As mentioned before, in deriving the master equation \eqref{finlmseqn} it has been assumed that the time-delay $\tau$ introduced by the feedback loop is negligible \cite{hmwi94}. In this section we explore a complementary regime when the feedback loop introduces a non-negligible time-delay $\tau$. In particular, we use  the time-delayed feedback control method of Ref.~\cite{kpyrg92,nyam14,algr14} and apply it to our Hamiltonian \eqref{lincpledosc}. We refer the reader to ~\cite{algr14} for a specific proposal implementing a finite time-delay in an optical feedback loop and applying it to the Dicke model \eqref{lincpledosc}. In this work we go beyond the analysis presented in \cite{algr14} and will include quantum fluctuations to check the steady state stability of the  time-delayed feedback control scheme and to compute steady state correlations between modes $\hat{a},\hat{b}$. Considering a specific all-optical time-delayed  feedback control strategy discussed in  detail in~\cite{algr14} and applying it to our  Hamiltonian  \eqref{lincpledosc}  we arrive at the following Heisenberg-Langevin equations of motion 
\eqref{lincpledosc} 
\begin{eqnarray}\label{eqnfedqm}
\frac{d}{dt} {\mbox V}(t)^{T} &=& {\mbox{\bf A}}{\mbox V}(t)^{T} - {\mbox{\bf B}}{\mbox V}(t)^{T}\nonumber \\
&& \displaystyle{-\sqrt{2\Gamma}\,{\mbox V}_{in}(t)^{T}+ {\mbox{\bf B}}{\mbox V}(t-\tau)^{T}},
\end{eqnarray}
where 
\begin{equation}
\begin{array}{l}
 \mbox V(t)=(\hat{a}(t),\hat{a}^{\dagger}(t),\hat{b}(t),\hat{b}^{\dagger}(t)),  \\ \\
 {\mbox V}(t-\tau)=(\hat{a}(t-\tau),\hat{a}^{\dagger}(t-\tau),\hat{b}(t-\tau),\hat{b}^{\dagger}(t-\tau)),  \\ \\
 {\mbox V}_{in}(t)=(\hat{a}^{{in}}(t),\hat{a}^{\dagger in}(t),0,0), \\ \\
 {\mbox{\bf A}}=\left( \begin{array}{cccc}
   -i\omega-\gamma & 0 & -i \lambda_1 &-i\lambda_2 \\
     0 & i\omega-\gamma & i \lambda_2 &i \lambda_1  \\
      -i \lambda_1 &-i\lambda_2 & -i\Omega&0 \\
     i \lambda_2 &i \lambda_1  &0 &i\Omega\\
  \end{array} \right ), \\ \\
   {\mbox{\bf B}}=\left( \begin{array}{cccc}
   \gamma/2 & 0 & 0 & 0 \\
     0 & \gamma/2 & 0 & 0  \\
      0 &0 & 0&0 \\
     0 &0 &0 &0\\
  \end{array} \right ).
 \end{array}
 \end{equation}
Here, $V_{in}(t)$ contains the input noise terms~\cite{walls}. It should be pointed out that in writing the above Heisenberg-Langevin equations we have assumed that the time-delayed feedback loop has unit efficiency~\cite{algr14}. To connect with the approach of the previous sections, we have again assumed that the damping of the mode $\hat{a}$  is the only dominant channel of dissipation. Also,  the time-delayed control feedback strategy \eqref{eqnfedqm} is implemented through a control force which is generated from the difference between the instantaneous cavity field $\hat{a}(t)$ and the field at some point in the past $\hat{a}(t-\tau)$ ~\cite{algr14}.

\begin{figure}[h!]
  \centering
    \includegraphics[width=0.5\textwidth]{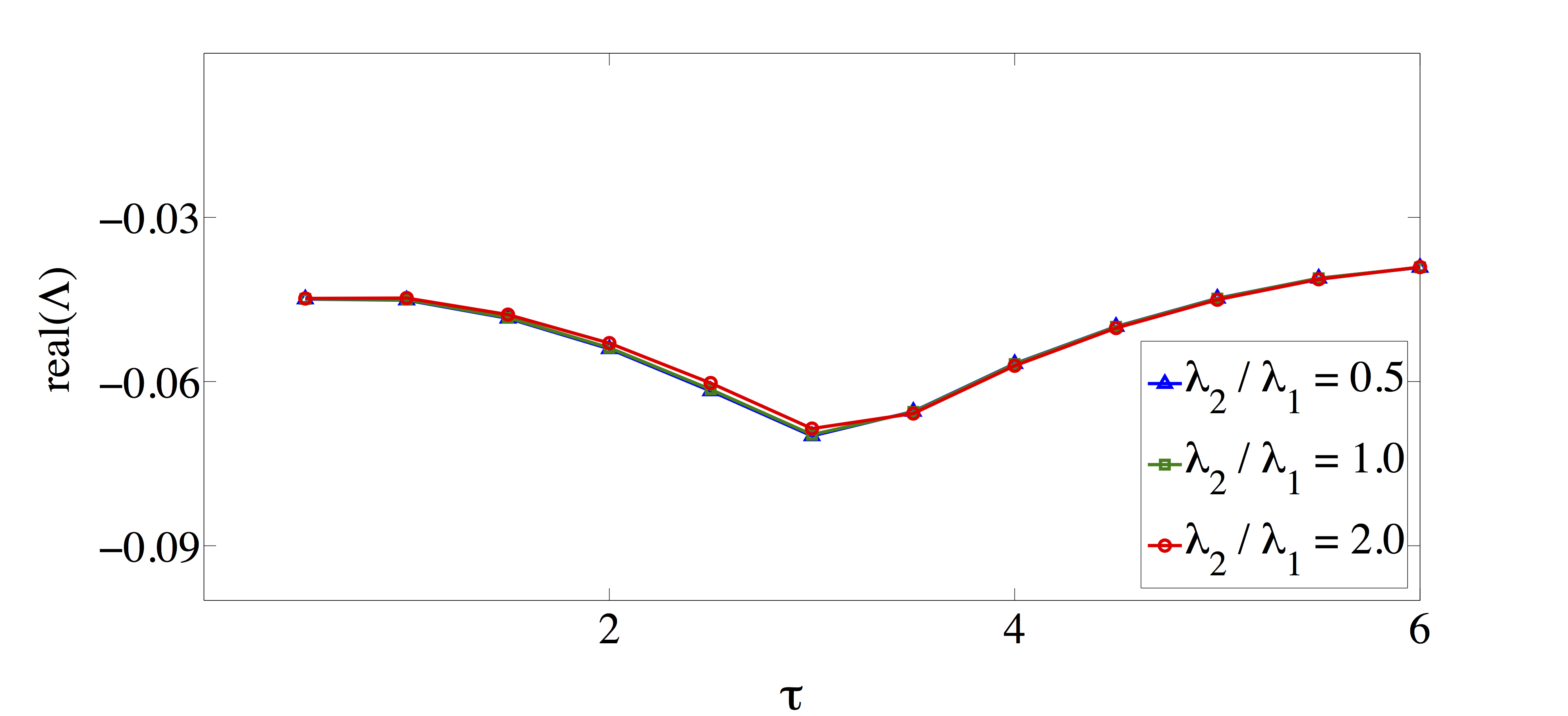}\\
        \caption{(Color online) Maximum of the real part of all possible solutions of the secular equation \eqref{creqntimdely} for three different values of the ratio $\lambda_{2}/\lambda_{1}$ and shown here as a function of the time-delay $\tau$. The dimensionless parameters have been taken $\Omega=\omega=1, \lambda_1=0.1$ and $\gamma=0.1$. For these set of parameters the steady state is stable for all values of the ratio $0 \leq \lambda_2/\lambda_1 \leq 2$ in the absence of a time-delayed feedback loop, see Fig.~\ref{stblinosc}.}
        \label{fidstblty}
\end{figure}

As a first step to check the influence of the time-delayed feedback control on our hybrid quantum system we check the stability of the above Heisenberg-Langevin equations semi-classically, {\it i.e.} we assume the noise $\langle  {\mbox V}_{in}(t) \rangle$=0. Using an ansatz $V(t) \sim e^{\Lambda t}$ we get the following secular equation~\cite{kpyrg92,nyam14} 
\begin{equation}\label{creqntimdely}
{\rm det}({\mbox{\bf A}}-{\mbox{\bf B}}+{\mbox{\bf B}}e^{-\Lambda \tau}-\Lambda \mathbb{1}).
\end{equation}
For a non-zero value of $\tau$, this transcendental equation has an infinite set of complex solutions for  eigenvalues $\Lambda$. The steady state is stable only if the real parts of all the solutions $\Lambda$ are negative~\cite{kpyrg92,nyam14}. We numerically solve the above secular equation \eqref{creqntimdely} for all possible roots $\Lambda$. In doing so we choose $\lambda_1$ and $\gamma$ such that in the absence of a time-delayed feedback loop the steady state is stable for all values of the ratio $ 0 \leq \lambda_2/\lambda_1 \leq 2$, see Fig.~\ref{stblinosc}. In Fig.~\ref{fidstblty} we plot the maximum of the real part of all possible solutions $\Lambda$. We find that at a semiclassical level, and for the set of parameters considered in Fig.~\ref{fidstblty}, our  time-delayed feedback control strategy does not qualitatively alter the stability of the steady state. 

\begin{figure}[h!]
  \centering
  \includegraphics[width=0.48\textwidth]{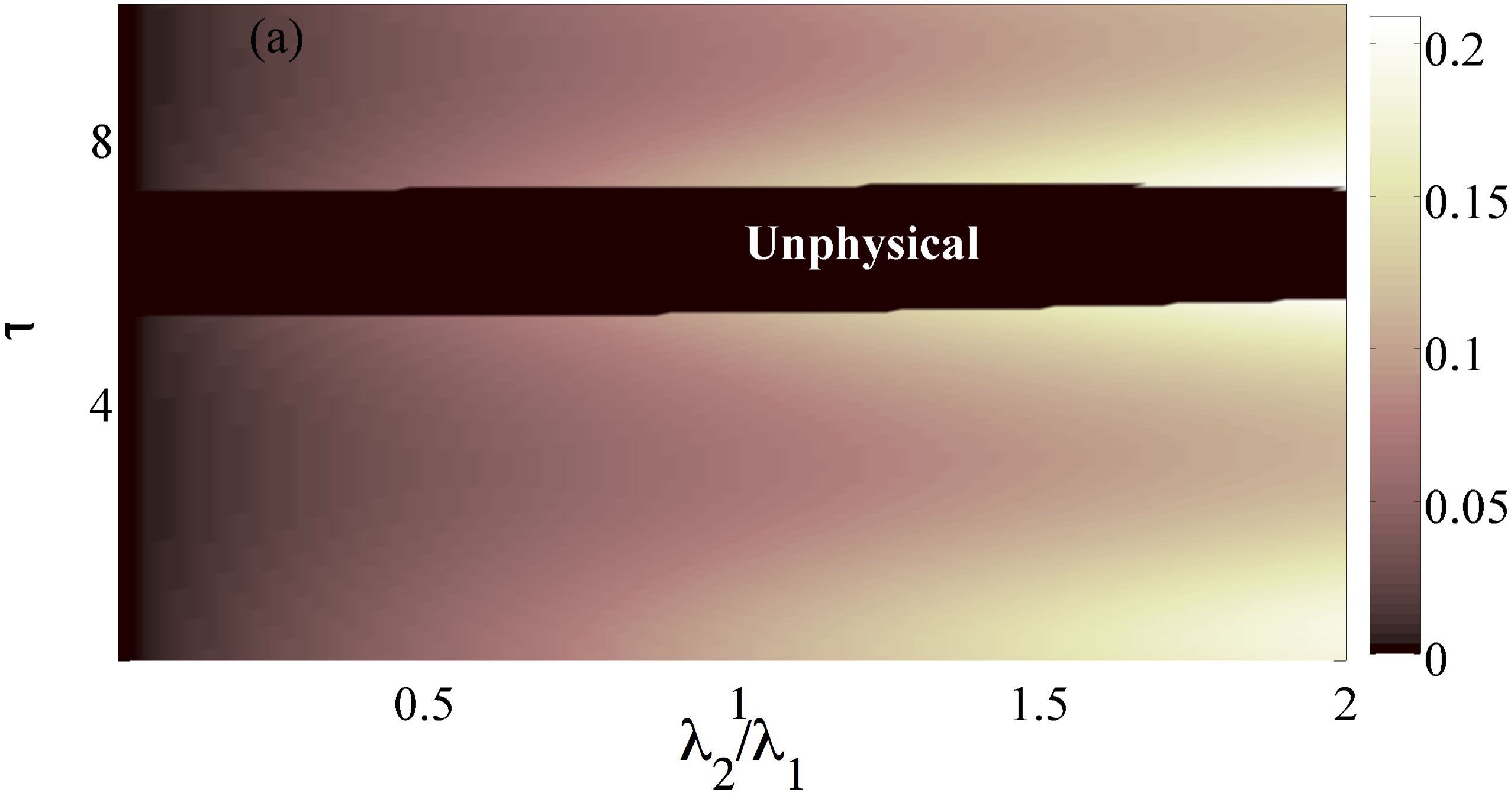}\\%
    \includegraphics[width=0.5\textwidth]{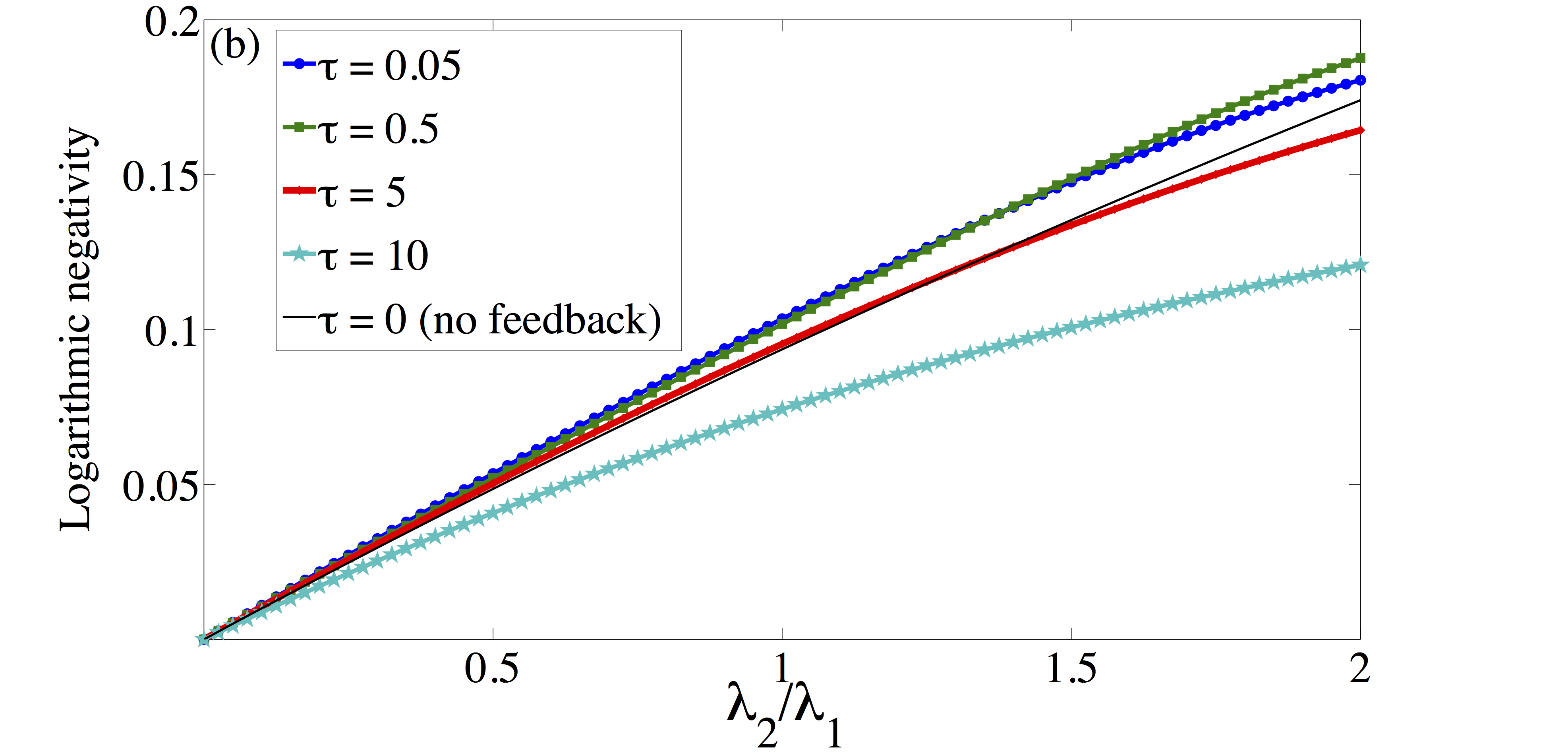}%
        \caption{(Color online)  
The upper plot (a) displays the bi-partite entanglement (negativity) between the modes $\hat{a}$ and $\hat{b}$ as a function of the dimensionless ratio $\lambda_{2}/\lambda_{1}$ and the time-delay $\tau$, and in the lower plot (b) the entanglement is given for different ratios $\lambda_2/\lambda_1$ and  for different choices  of $\tau$. Also shown in (b) is the steady state bi-partite entanglement between the modes $\hat{a}$ and $\hat{b}$ in the absence of a time-delayed feedback loop, and we especially note that by introducing a time-delay the amount of entanglement can be increased. In the black region in (a) the solution does not approach a physical steady state. The dimensionless parameters have been taken $\Omega=\omega=1, \lambda_1=0.1$ and $\gamma=0.1$.}
\label{entwithtimedelydfed}
\end{figure}

To show that our time-delayed feedback control strategy is indeed capable of influencing the  steady state behaviour of our hybrid quantum system, we now embark on a full quantum treatment to explore the steady state correlations between the modes $\hat a$ and $\hat b$ with their dynamics governed by Eq.~\eqref{eqnfedqm}.  We solve the equations  in the Fourier space and reconstruct the steady state two-mode covariance matrix in order to capture the quantum correlations between the modes $\hat{a}$ and $\hat{b}$. We then evaluate the  bi-partite entanglement (logarithmic negativity) between the two modes $\hat{a}$ and $\hat{b}$. In Fig.~\ref{entwithtimedelydfed} (a) we display the bi-partite entanglement between the modes $\hat{a}$ and $\hat{b}$ as a function of the dimensionless ratio $\lambda_2/\lambda_1$ and the time-delay $\tau$. As for Fig.~\ref{fidstblty}, we choose $\lambda_1$ and $\gamma$ such that in the absence of a time-delayed feedback loop the steady state is stable for all values of the ratio $ 0 \leq \lambda_2/\lambda_1 \leq 2$. We find that for a fixed value of the time-delay $\tau$ the bi-partite entanglement between the modes $\hat{a}$ and $\hat{b}$ increases with  the value of the ratio $\lambda_2/ \lambda_1$, this is expected since  the coupling between the two modes is also increased. However,  we also observe that there is a ``delay-window'' (shown in black and labeled `Unphysical' in Fig.~\ref{entwithtimedelydfed} (a)), such that if $\tau$ is chosen within this interval then  a physical steady state is never reached \cite{commentonstab}. We also find that a suitable choice of the time-delay can also increase the steady state bi-partite entanglement between the modes $\hat{a}$ and $\hat{b}$. More precisely, the bi-partite entanglement between the modes $\hat{a}$ and $\hat{b}$ for different choices of $\tau$ is given in Fig.~\ref{entwithtimedelydfed} (b), where for the sake of elucidating the role of time-delayed feedback loop we have also shown the steady state entanglement between the modes $\hat{a}$ and $\hat{b}$ in the absence of a time-delayed feedback loop.

\section{Conclusions}
\label {sec:concld}
In this work we have studied two topical hybrid systems, the anisotropic Rabi and Dicke models, both of which combine disparate quantum degrees of freedom; light and matter. While similar, the two models operate in different `regimes'; the Rabi model displays large quantum fluctuations which is not the case of the Dicke model. We have first explored the open dynamics of a two-level system (a model atom) strongly coupled to a boson (photon) field, i.e. the Rabi model. We have shown that the NESS is an entangled state of the light and atom systems. More precisely, the steady state is a statistical mixture of two states with opposite parities. In phase space these even and odd parity states forms a cat-like structure, which however gets `tilted' due to the reservoir-induced Lamb shift. As a second physical model we have investigated two coupled boson modes which can describe linearized interactions between a cavity field and a collection of two-level systems in the thermodynamic limit, i.e. the Dicke model. This linearized version of the anisotropic Dicke model describes the collective excitations in the normal phase. We have especially explored the stability and bi-partite entanglement in this second physical setting, again in the presence of coupling to an environment. Even though quantum fluctuations are negligible in the thermodynamic limit, entanglement survives which we attribute to the presence of a critical point which persists also under dissipation. As a way to establish quantum control and increase non-classical properties in such hybrid quantum architectures, we propose to use an all optical feedback strategy. We demonstrated this approach by utilizing the scheme for the Dicke model, for two applications. We have applied feedback to provide protection (against loss) of quantum states of interest for metrology (NOON and entangled coherent states).  We have also demonstrated that  a realistic feedback  proposal including a time-delay can be used to alter the properties of the NESS potentially increasing the entanglement between the sub-systems and changing the stability of the steady state.

\acknowledgments{}
C. J. acknowledges the York Centre for Quantum Technologies (YCQT) Fellowship. J. L. acknowledges VR-Vetenskapsr\aa det (The Swedish Research Council) and KAW (The Knut
and Alice Wallenberg foundation) for financial support. We gratefully acknowledge useful discussions with Gerard Milburn and Jaewoo Joo. 

\appendix
\section{Markovian master equation with optical feedback}\label{appendix1}
In this appendix we provide the details of the derivation of the master equation \eqref{finlmseqn}. For the feedback loop to be effective, the driven cavity should respond much faster than the source cavity. We thus work in a regime where $\gamma_\mathrm{d}\gg1$  and the state of the driven cavity follows adiabatically the evolution of the source cavity. The large decay rate and the coupling to a zero temperature reservoir implies that driven cavity may only be weakly excited, which justifies the ansatz state (\ref{appstate}). With such a joint state of two coupled cavities one obtains the following equations of motion for the source cavity 
  \begin{eqnarray}
\dot{\hat{\rho}}_{00}&=&-i\left[\hat H_{\rm nAD},\hat{\rho}_{00}\right]+\frac{\gamma}{2} \mathcal L_{\hat{a}}\hat{\rho}_{00}+\gamma_\mathrm{d} \hat{\rho}_{11}\nonumber \\
&&+\sqrt{\gamma\gamma_\mathrm{d}}(\hat{a}\hat{\rho}_{10}^{\dagger}+\hat{\rho}_{10}\hat{a}^{\dagger}) +\mu\frac{\sqrt{\gamma\gamma_\mathrm{d}}}{2}(\hat{a}^{\dagger}\hat{\rho}_{10}+\hat{\rho}_{10}^{\dagger}\hat{a})\nonumber \\
\dot{\hat{\rho}}_{10}&=&-i\left[\hat H_{\rm nAD},\hat{\rho}_{10}\right]+\frac{\gamma}{2}  \mathcal L_{\hat{a}}\hat{\rho}_{10}-\frac{\gamma_\mathrm{d}}{2} \hat{\rho}_{10}\nonumber \\
&&+\sqrt{\gamma\gamma_\mathrm{d}}(\hat{a}\hat{\rho}_{11}-\hat{a}\hat{\rho}_{00}+\sqrt{2}\hat{\rho}_{20}\hat{a}^{\dagger})\nonumber \\
&&+\mu\frac{\sqrt{\gamma\gamma_\mathrm{d}}}{2}(\sqrt{2}\hat{a}^{\dagger}\hat{\rho}_{20}-\hat{a}\hat{\rho}_{00}+\hat{\rho}_{11}\hat{a})\nonumber \\
\dot{\hat{\rho}}_{11}&=&-i\left[\hat H_{\rm nAD},\hat{\rho}_{11}\right]+\frac{\gamma}{2} \mathcal L_{\hat{a}}\hat{\rho}_{11}-(i\Omega_\mathrm{d}+\gamma_\mathrm{d})\hat{\rho}_{11}\nonumber \\
&&-\sqrt{\gamma\gamma_\mathrm{d}}(\hat{a}\hat{\rho}_{10}^{\dagger}+\hat{\rho}_{10}\hat{a}^{\dagger})\nonumber -\mu\frac{\sqrt{\gamma\gamma_\mathrm{d}}}{2}(\hat{a}\hat{\rho}_{10}^{\dagger}+\hat{\rho}_{10}\hat{a}^{\dagger})\nonumber \\
\dot{\hat{\rho}}_{20}&=&-i\left[\hat H_{\rm nAD},\hat{\rho}_{20}\right]+\frac{\gamma}{2} \mathcal L_{\hat{a}}\hat{\rho}_{20}-2\gamma_\mathrm{d} \hat{\rho}_{20}\nonumber \\
&&-\sqrt{2\gamma\gamma_\mathrm{d}}\hat{a}\hat{\rho}_{10}-\mu\frac{\sqrt{\gamma\gamma_\mathrm{d}}}{\sqrt{2}}\hat{a}\hat{\rho}_{10}\nonumber.
\end{eqnarray}

The state of the source cavity is of interest to us and it can be  extracted as $\hat{\rho}=\rm{Tr}_{\hat{c}}\hat{W}=\hat{\rho}_{00}+\hat{\rho}_{11}$. From the above equations, the off-diagonal elements $\hat{\rho}_{10}$ and $\hat{\rho}_{20}$ can be adiabatically eliminated by slaving them to the diagonal elements $\hat{\rho}_{00}$ and $\hat{\rho}_{11}$. Setting $\dot{\hat{\rho}}_{20}$=0 we obtain to leading order in $1/\sqrt{(\gamma_\mathrm{d}/\gamma)}$
\begin{equation} 
\hat{\rho}_{20}=-\frac{1}{\sqrt{2(\gamma_\mathrm{d}/\gamma)}}(\frac{\mu}{2}+1)\,\hat{a}\,\hat{\rho}_{10}.
\end{equation}
Inserting the above expression in the steady state equation of $\hat{\rho}_{10}$ we obtain
\begin{equation}
\hat{\rho}_{10}=\frac{2}{\sqrt{(\gamma_\mathrm{d}/\gamma)}}(\frac{\mu}{2}\hat{\rho}_{11}\hat{a}-\frac{\mu}{2}\hat{a}\hat{\rho}_{00}+\hat{a}\hat{\rho}_{11}-\hat{a}\hat{\rho}_{00}).
\end{equation}
Using the above steady state solution for the off-diagonal element yields  the following master equation for the density matrix of the source cavity alone
\begin{eqnarray}
\dot{\hat{\rho}}&=&\dot{\hat{\rho}}_{00}+\dot{\hat{\rho}}_{11}=
 -i[\hat H_{\rm nAD},\hat{\rho}]+\frac{\gamma}{2} \mathcal L_{\hat{a}}\hat{\rho} -i\Omega_\mathrm{d}\hat{\rho}_{11}\nonumber \\
&&+\mu \gamma[\mathcal L_{\hat{a}}(\hat{\rho}_{00}-\hat{\rho}_{11})+\frac{\mu}{2}\mathcal L_{\hat{a}}\hat{\rho}_{00}+\frac{\mu}{2}\mathcal L_{\hat{a}^{\dagger}}\hat{\rho}_{11}].
\end{eqnarray}
When $\gamma_\mathrm{d}\gg1$,  $\hat{\rho}_{11} \sim \mathcal O(0)$ and  $\hat{\rho} \approx \hat{\rho}_{00}$ we arrive at the following master equation approximating the source dynamics 
\begin{eqnarray}\label{largegammamseqn}
\dot{\hat{\rho}}&=&\dot{\hat{\rho}}_{00}+\dot{\hat{\rho}}_{11}=
 -i\left[\hat H_{\rm nAD},\hat{\rho}\right]+\frac{\gamma_{\rm eff}}{2}\mathcal L_{\hat{a}}\hat{\rho},
 \end{eqnarray}
 where $\gamma_{\rm eff}=\gamma(1+\mu(2+\mu))$ is the effective damping rate of the field mode in the source cavity. 
 
 \section{Uhlmann fidelity in the RWA regime}\label{appendix2}
 A conspicuous feature of Fig.~\ref{fidtimevolution} (especially in (c)) is the almost constant evolution of the fidelity for the states with $N=1,\,2$, and the oscillatory structure for the $N=3,\,4$ states. This implies that  $|\Psi_{-} ^{1,2}\rangle$ are stationary states while $|\Psi_{-} ^{3,4}\rangle$ seems to be formed from two stationary states. In this appendix we will explain how this comes about given the initial states and the Hamiltonian \eqref{lincpledosc}. As pointed out in the main text, for the figure a small coupling has been used (20 times smaller than the bare frequencies) which means that imposing the RWA is justified, i.e. we let $\lambda_2=0$ from now on.  We have numerically verified that the RWA is applicable for the corresponding parameters. 
 
Now, when $\lambda_{2}=0$ the Hamiltonian can be readily diagonalized by defining two new bosonic operators; `even'  $\hat{z}_{+}=(\hat{a}+ \hat{b})/\sqrt{2}$ and `odd' $\hat{z}_{-}=(\hat{a}- \hat{b})/\sqrt{2}$. These obey the regular boson commutation relations and mutually commute. 
In particular the `odd' Fock states $|n_-\rangle=\hat z_-^{\dagger n_-}|0\rangle/\sqrt{n_-!}$ have shown to be important for adiabatic passage in multimode cavities~\cite{erika}. Expressed with the new rotated operators, the Hamiltonian is diagonal
\begin{equation}
H_{z}=(\omega +\lambda_{1})\hat{z}^{\dagger}_{+}\hat{z}_{+}+(\omega -\lambda_{1})\hat{z}^{\dagger}_{-}\hat{z}_{-}.
\end{equation}
It follows that the eigenstates are $|n_+\rangle|n_-\rangle=\hat z_+^{\dagger n_+}\hat z_-^{\dagger n_-}|0\rangle_a|0\rangle_b/\sqrt{n_+!n_-!}$, with corresponding eigenenergies $\varepsilon_{n_+n_-}=\omega\left(n_++n_-\right)+\lambda_1\left(n_+-n_-\right)$.

For an $N$ particle NOON state we have
\begin{equation}
|\Psi_{-} ^{N}\rangle=\frac{1}{\sqrt{2N!}}\left(\hat a^{\dagger N}-\hat b^{\dagger N}\right)|0\rangle_a|0\rangle_b.
\end{equation}
Thus, we notice that $|\Psi_-^1\rangle=\hat z_-^\dagger|0\rangle_a|0\rangle_b$ is indeed a stationary state with its time evolution given by
\begin{equation}\label{apprxn1}
|\Psi_{-}^{1}(t)\rangle =e^{-i(\omega-\lambda_1) t}|\Psi_-^1 \rangle.
\end{equation}
The time evolved state \eqref{apprxn1} is of course strictly only correct within the RWA, and the exact time evolved state should be given by evolution under the full Hamiltonian; $|\Psi_{\rm exact}^{1}(t) \rangle=e^{-i\hat H_{\rm nAD} t}|\Psi_{-}^{1}\rangle$. Numerically we find the error arriving from neglecting the counter rotating terms $\delta_{1}=|\Phi_{\rm exact}-\Phi_{\rm RWA}| < 0.003$ for all $t$. 
Similarly, we find that the two particle NOON state $|\Psi_-^2\rangle=\hat z_+^\dagger\hat z_-^\dagger|0\rangle_a|0\rangle_b$ is also a stationary state with the time evolution
\begin{equation}\label{apprxn2}
|\Psi_{-}^{2}(t)\rangle =e^{-i 2\omega t}|\Psi_-^2 \rangle.
\end{equation}
In this case the numerically estimated error $\delta_{2}=|\Phi_{\rm exact}-\Phi_{\rm approx}| < 0.004$ for all $t$. Interestingly, the alternative NOON state $|\Psi_{+} ^{2}\rangle =(|2\rangle_{a} |0 \rangle_{b}+|0 \rangle |_{a}2 \rangle_{b} )/\sqrt{2}$ is not a stationary state. 

Turning now to the $N=3$ case. It is easy to show that there is no $N=3$ NOON state (most general form $|3\rangle_a|0\rangle_b+e^{i\phi}|0\rangle_a|3\rangle_b$) that can be a stationary state. For the present state the time evolved state takes the form
 \begin{equation}\label{apprxn34othgnl}
|\Psi_{-}^{3}(t)\rangle=\frac{e^{-i 3\omega t}}{2\sqrt{3!}}(3\hat{z}_{+}^{\dagger 2}\hat{z}_{-}^{\dagger}e^{-i \lambda_{1} t}+\hat{z}_{-}^{\dagger 3}e^{i 3\lambda_{1} t})|0\rangle_a|0\rangle_b.
\end{equation}
It is a straightforward exercise to show that $|\Psi_{-}^{3}(t)\rangle$ evolves as a time dependent mixture of the orthogonal states $|\Psi_{-}^{3}\rangle$ and  $|\Psi_{\rm orth} ^{3}\rangle =(|2\rangle_{a} |1 \rangle_{b}-|1 \rangle_{a}|2 \rangle_{b} )/\sqrt{2}$. But, never in the course of time evolution the coefficient of  $|\Psi_{-}^{3}\rangle$ vanishes completely. This is the reason why the fidelity $\Phi$ for $N=3$ in Fig.~\ref{fidtimevolution} (c) never drops close to zero. 

The situation is however somewhat different for $|\Psi_{-}^{4}(t)\rangle$ whose time evolution is
\begin{equation}
|\Psi_{-}^{4}(t)\rangle =\frac{e^{-i4\omega t}\sqrt{2}}{\sqrt{4!}}(\hat{z}_{+}^{\dagger 3}\hat{z}_{-}^{\dagger}e^{-i 2\lambda_{1} t}+\hat{z}_{+}^{\dagger}\hat{z}_{-}^{\dagger 3}e^{i 2\lambda_{1} t})|0\rangle_a|0\rangle_b. 
\end{equation}
Even though, like in the $N=3$ situation, one cannot find a general NOON state being a stationary state (in fact it is only possible for $N=1,\,2$), the above time evolved state evolves into a state completely orthogonal to $|\Psi_{-}^{4}\rangle$. For instance, when  $2\lambda_{1}t=(2m+1) \pi/2$ one obtains $|\Psi_{-}^{4}(t)\rangle \sim |\Psi_{\rm orth} ^{4}\rangle =(|3\rangle_{a} |1 \rangle_{b}-|1 \rangle_{a}|3 \rangle_{b} )/\sqrt{2}$. 
This is where  the fidelity $\Phi$ for  $|\Psi_{-}^{4}\rangle$ drops close to zero as shown in Fig.~\ref{fidtimevolution} (c). 
 


\begin{thebibliography}{99}
\bibitem{commntdoma2} See Ref. \cite{doma2}. We do, however, also refer the reader to \cite{jspe15} for a recent proposal on simulating the ultra-strong and deep strong coupling regimes of quantum Rabi model with trapped ions.
\bibitem{commentonstab}  We demand all  physical steady states should satisfy canonical commutation relations. This constraint  can be conveniently formulated in terms of symplectic eigenvalues for Gaussian states, see  \cite{gera07} for details.
\end{thebibliography}
\end{document}